\documentclass{nature}

\usepackage{graphicx}
\usepackage{xcolor}
\usepackage{amsmath}	
\usepackage{amssymb}
\usepackage{mathtools}
\usepackage{subcaption}
\usepackage{textcomp}
\usepackage{gensymb}
\usepackage{hyperref}
\usepackage{enumitem}

\usepackage{lineno} 
\usepackage[normalem]{ulem}

\usepackage{etoolbox}

\title{HAWC observations of the acceleration of very-high-energy cosmic rays in the Cygnus Cocoon}

\newcounter{firstbib}

\begin{document}

\author{{A.U.~Abeysekara}$^1$, {A.~Albert}$^2$, {R.~Alfaro}$^3$,
{C.~Alvarez}$^4$, {J.R.~Angeles Camacho}$^3$, {J.C.~Arteaga-Velázquez}$^5$, {K.P.~Arunbabu}$^6$, {D.~Avila Rojas}$^3$, {H.A.~Ayala Solares}$^7$, {V.~Baghmanyan}$^8$, {E.~Belmont-Moreno}$^5$, {S.Y.~BenZvi}$^9$, {R.~Blandford}\mbox{*}$^{10}$, {C.~Brisbois}$^{11}$, {K.S.~Caballero-Mora}$^4$, {T.~Capistrán}$^{12, 13}$, {A.~Carramiñana}$^{12}$, {S.~Casanova}\mbox{*}$^8$, {U.~Cotti}$^5$, {S.~Coutiño de León}$^{12}$, {E.~De la Fuente}$^{14, 15}$, {R.~Diaz Hernandez}$^{12}$, {B.L.~Dingus}$^2$, {M.A.~DuVernois}$^{16}$, {M.~Durocher}$^2$, {J.C.~Díaz-Vélez}$^{14}$, {R.W.~Ellsworth}$^{11}$, {K.~Engel}$^{11}$, {C.~Espinoza}$^3$, {K.L.~Fan}$^{11}$, {K.~Fang}\mbox{*}$^{10, 16}$, {H.~Fleischhack}\mbox{*}$^{17}$, {N.~Fraija}$^{13}$, {A.~Galván-Gámez}$^{13}$, {D.~Garcia}$^3$, {J.A.~García-González}$^3$, {F.~Garfias}$^{13}$, {G.~Giacinti}$^{18}$, {M.M.~González}$^{13}$, {J.A.~Goodman}$^{11}$, {J.P.~Harding}$^2$, {S.~Hernandez}$^3$, {J.~Hinton}$^{18}$, {B.~Hona}$\mbox{*}^1$, {D.~Huang}$^{17}$, {F.~Hueyotl-Zahuantitla}$^4$, {P.~Hüntemeyer}\mbox{*}$^{17}$, {A.~Iriarte}$^{13}$, {A.~Jardin-Blicq}$^{18, 19, 20}$, {V.~Joshi}$^{21}$, {D.~Kieda}$^1$, {A.~Lara}$^6$, {W.H.~Lee}$^{13}$, {H.~León Vargas}$^3$, {J.T.~Linnemann}$^{22}$, {A.L.~Longinotti}$^{12, 13}$, {G.~Luis-Raya}$^{23}$, {J.~Lundeen}$^{12}$, {K.~Malone}$^2$, {O.~Martinez}$^{24}$, {I.~Martinez-Castellanos}$^{11}$, {J.~Martínez-Castro}$^{25}$, {J.A.~Matthews}$^{26}$, {P.~Miranda-Romagnoli}$^{27}$, {J.A.~Morales-Soto}$^5$, {E.~Moreno}$^{24}$, {M.~Mostafá}$^7$, {A.~Nayerhoda}$^8$, {L.~Nellen}$^{28}$, {M.~Newbold}$^1$, {M.U.~Nisa}$^{22}$, {R.~Noriega-Papaqui}$^{27}$, {L.~Olivera-Nieto}$^{18}$, {N.~Omodei}$^{10}$, {A.~Peisker}$^{22}$, {Y.~Pérez Araujo}$^{13}$, {E.G.~Pérez-Pérez}$^{23}$, {Z.~Ren}$^{26}$, {C.D.~Rho}$^{29}$, {D.~Rosa-González}$^{12}$, {E.~Ruiz-Velasco}$^{18}$, {H.~Salazar}$^{24}$, {F.~Salesa Greus}$^{8, 30}$, {A.~Sandoval}$^3$, {M.~Schneider}$^{11}$, {H.~Schoorlemmer}$^{18}$, {F.~Serna}$^3$, {A.J.~Smith}$^{11}$, {R.W.~Springer}$^1$, {P.~Surajbali}$^{18}$, {K.~Tollefson}$^{22}$, {I.~Torres}$^{12}$, {R.~Torres-Escobedo}$^{14}$, {F.~Ureña-Mena}$^{12}$, {T.~Weisgarber}$^{31}$, {F.~Werner}$^{18}$, {E.~Willox}$^{11}$, {A.~Zepeda}$^{32}$, {H.~Zhou}$^{33}$, {C.~De León}$^5$  \& {J.D.~Álvarez}$^5$}

\maketitle
\begin{affiliations}
\item {Department of Physics and Astronomy, University of Utah, Salt Lake City, UT, USA }

\item {Physics Division, Los Alamos National Laboratory, Los Alamos, NM, USA }

\item {Instituto de F\'{i}sica, Universidad Nacional Autónoma de México, Ciudad de Mexico, Mexico }

\item {Universidad Autónoma de Chiapas, Tuxtla Gutiérrez, Chiapas, México}

\item {Universidad Michoacana de San Nicolás de Hidalgo, Morelia, Mexico }

\item {Instituto de Geof\'{i}sica, Universidad Nacional Autónoma de México, Ciudad de Mexico, Mexico }

\item {Department of Physics, Pennsylvania State University, University Park, PA, USA }

\item {Institute of Nuclear Physics Polish Academy of Sciences, PL-31342 IFJ-PAN, Krakow, Poland }

\item {Department of Physics \& Astronomy, University of Rochester, Rochester, NY , USA }

\item {Department of Physics, Stanford University: Stanford, CA 94305–4060, USA}

\item {Department of Physics, University of Maryland, College Park, MD, USA }

\item {Instituto Nacional de Astrof\'{i}sica, Óptica y Electrónica, Puebla, Mexico }

\item {Instituto de Astronom\'{i}a, Universidad Nacional Autónoma de México, Ciudad de Mexico, Mexico }

\item {Departamento de F\'{i}sica, Centro Universitario de Ciencias Exactase Ingenierias, Universidad de Guadalajara, Guadalajara, Mexico }

\item{Institute for Cosmic Ray Research, University of Tokyo: 277-8582 Chiba, Kashiwa, Kashiwanoha, 5 Chome-1-5}

\item {Department of Physics, University of Wisconsin-Madison, Madison, WI, USA }

\item {Department of Physics, Michigan Technological University, Houghton, MI, USA }

\item {Max-Planck Institute for Nuclear Physics, 69117 Heidelberg, Germany}

\item {Department of Physics, Faculty of Science, Chulalongkorn University, 254
Phayathai Road, Pathumwan, Bangkok 10330, Thailand}

\item{National Astronomical Research Institute of Thailand (Public
Organization), Don Kaeo, MaeRim, Chiang Mai 50180, Thailand}

\item {Erlangen Centre for Astroparticle Physics, Friedrich-Alexander-Universit\"at Erlangen-N\"urnberg, Erlangen, Germany}

\item {Department of Physics and Astronomy, Michigan State University, East Lansing, MI, USA }

\item {Universidad Politecnica de Pachuca, Pachuca, Hgo, Mexico }

\item {Facultad de Ciencias F\'{i}sico Matemáticas, Benemérita Universidad Autónoma de Puebla, Puebla, Mexico }

\item {Centro de Investigaci\'on en Computaci\'on, Instituto Polit\'ecnico Nacional, M\'exico City, M\'exico.}

\item {Dept of Physics and Astronomy, University of New Mexico, Albuquerque, NM, USA }

\item {Universidad Autónoma del Estado de Hidalgo, Pachuca, Mexico }

\item {Instituto de Ciencias Nucleares, Universidad Nacional Autónoma de Mexico, Ciudad de Mexico, Mexico }

\item {Natural Science Research Institute, University of Seoul, Seoul, Republic Of Korea}

\item{Instituto de F\'isica Corpuscular, CSIC, Universitat de Val\`encia, E-46980, Paterna, Valencia, Spain}

\item {Department of Chemistry and Physics, California University of Pennsylvania, California, Pennsylvania, USA}

\item {Physics Department, Centro de Investigacion y de Estudios Avanzados del IPN, Mexico City, DF, Mexico }

\item{Tsung-Dao Lee Institute \&{} School of Physics and Astronomy, Shanghai Jiao Tong University, Shanghai, China }

\end{affiliations}

\begin{abstract}

Cosmic rays  with energies up to a few PeV are known to be accelerated within the Milky Way \cite{hillasbook, 1990acr..book.....B}. Traditionally, it has been presumed that supernova remnants were the main source of very-high-energy cosmic rays \cite{Baade259,  H_randel_2004} but theoretically it is difficult to get protons to PeV energies \cite{Bell:2013kq, 2013APh....43...71A} and  observationally there simply is no evidence to support the remnants as sources of hadrons with energies above a few tens of TeV \cite{2012SSRv..173..369H, 2017hsn..book.1737F}. One possible source of protons with those energies is the Galactic Center region \cite{2016Natur.531..476H}. Here we report observations of 1-100 TeV $\gamma$ rays coming from the ‘Cygnus Cocoon’ \cite{fermicocoon}, which is a superbubble surrounding a region of OB2 massive star formation. These $\gamma$ rays are likely produced by 10-1000 TeV freshly accelerated CRs originating from the enclosed star forming region Cygnus OB2. Hitherto it was not known that such regions could accelerate particles to these energies. The measured flux is likely originated by hadronic interactions. The spectral shape and the emission profile of the Cocoon changes from GeV to TeV energies, which reveals the transport of cosmic particles and historical activity in the superbubble.
\end{abstract}

The High Altitude Water Cherenkov (HAWC) observatory is a wide field-of-view very-high-eenergy (VHE) $\gamma$-ray instrument sensitive in the energy range of 300~GeV to beyond 100~TeV. It is uniquely suited to study extended emission regions that contain bright background sources, as is the case for the Cygnus superbubble. A bright source, named 2HWC~J2031+415 in the second HAWC catalog \cite{catalog} as shown in the significance map in Fig. 1, has been detected coincident with the superbubble. The location of this $\gamma$-ray emission overlaps with a known pulsar wind nebula (PWN) TeV J2032+4130 \cite{pwn2002}. Both TeV J2032+4130 and 2HWC J2031+415 are situated well within the extended region of $\gamma$-ray emission detected at GeV energies by {\it Fermi}-LAT \cite{fermicocoon}. Another source, 2HWC J2020+403, possibly associated with the $\gamma$ Cygni SNR, lies 2.36\degree \ away from the center of the 2HWC J2031+415. 

\begin{figure}
\includegraphics[width=0.99\textwidth]{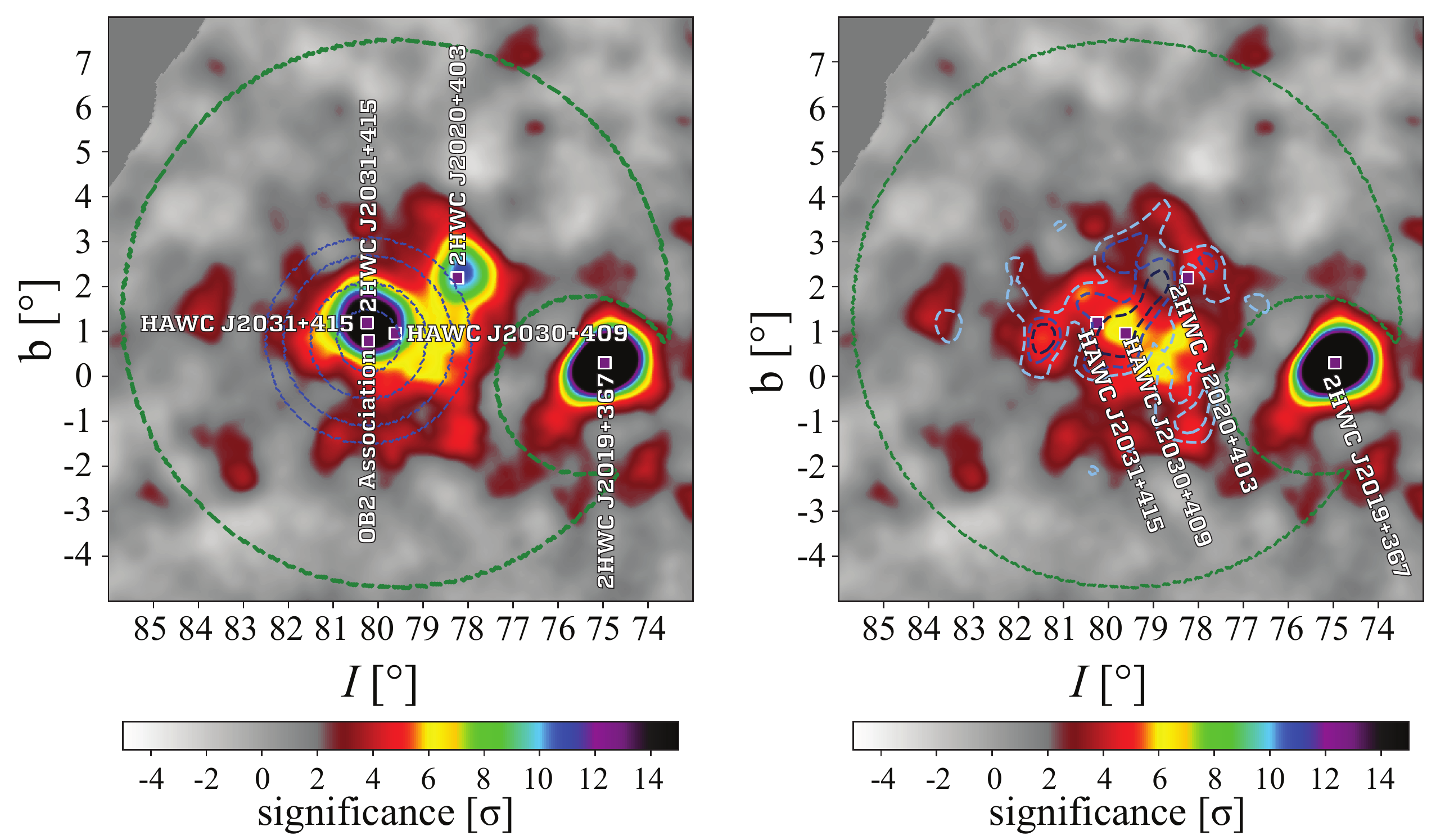}
\caption{\textbf{Significance map of the Cocoon region before and after subtraction of the known sources at the region.} \textbf{Left}: Significance map of the Cocoon region. The significance map is in Galactic coordinates where b and l refers to latitude and longitude. It is produced as described in \cite{catalog} with 0.5\degree{} extended disk source assumption and a spectral index of -2.6 using 1343 days of HAWC data. The blue contours are four annuli centered at the OB2 association  as listed in Supplementary Table 2. The green contour is the region of interest (ROI) used for the study which masks the bright source 2HWC J2019+367. \textbf{Right}: Significance map of the Cocoon region after subtracting HAWC J2031+415 (PWN) and 2HWC J2020+403 ($\gamma$ Cygni). The map is made assuming a 0.5\degree{} extended disk source and a spectral index of -2.6 with 1343 days of HAWC data. The lighter and darker blue dashed lines are 0.16, 0.24 and 0.32  photons/(0.1\degree{} by 0.1\degree{} spatial bin) contours from {\it Fermi}-LAT Cocoon \cite{fermicocoon}.}
\end{figure}

Using 1343 days of measurements with HAWC, we successfully removed the contribution of the overlapping sources to the TeV $\gamma$-ray emission in the ROI shown in Fig. 1. The 2HWC J2031+415 emission is well described by two sources: HAWC J2031+415 (at RA = 307.90\degree \ $\pm$ 0.04\degree, Dec = 41.51\degree \ $\pm$ 0.04\degree), a slightly extended source with a Gaussian width of 0.27\degree, possibly associated with the PWN TeV J2032+4130 \cite{pwn2002, pwn2005}, and HAWC J2030+409, a VHE counterpart of the GeV Cygnus Cocoon \cite{fermicocoon} (see Methods). The region after subtraction of HAWC J2031+415 (PWN) and 2HWC J2020+403 ($\gamma$ Cygni) is shown in Fig. 1.

HAWC J2030+409, contributes $\sim$  90\% to the total flux detected at the ROI and is detected with test statistics, TS (likelihood ratio test), of 195.2 (see Equation \ref{eq:TS}) at the position (RA = 307.65\degree \ $\pm$ 0.30\degree, Dec = 40.93\degree \ $\pm$ 0.26\degree). The extension is well described by a Gaussian profile with width of 2.13\degree$ \ \pm \ {0.15}\degree(\rm stat.) \pm \ 0.06\degree (\rm syst.)$. The location and the Gaussian width of the source are consistent with the measurements by {\it Fermi}-LAT above 1 GeV to a few hundred GeV.

The spectral energy distribution of the Cygnus Cocoon has been extended from 10~TeV in the previously published measurement by the ARGO observatory \cite{Bartoli14} to 200~TeV in this analysis. The measurement above 0.75~TeV can be described by a power-law spectrum $dN/dE=N_0\,(E/E_0)^\Gamma$, with $E_{0} = 4.2$~TeV being the pivot energy. The flux normalization is $N_{0}=9.3_{-0.8}^{+0.9} (\rm stat.) _{-1.23}^{+0.93} (\rm syst.)  \times$ $10^{-13}  \rm \, \,cm^{-2}\,s^{-1}\,TeV^{-1}$ and the spectral index is $\Gamma = -2.64_{-0.05}^{+0.05} (\rm stat.) _{-0.03}^{+0.09} (\rm syst.)$. The flux is compatible with an extrapolation from the {\it Fermi}-LAT measurement at 1-300~GeV \cite{fermicocoon, 4fgl}. Compared to $\Gamma = -2.1$ in the {\it Fermi}-LAT GeV data, a significant softening of the energy spectral density is evident at a few TeV in the ARGO data \cite{Bartoli14}, and persists beyond 100~TeV in the HAWC data [Fig. 2]. 

GeV $\gamma$ rays observed by {\it Fermi}-LAT can be produced either by high-energy protons interacting with gas or by high-energy electrons upscattering stellar radiation and dust emission \cite{fermicocoon}. Above a few TeV, the inverse-Compton process between relativistic electrons and stellar photons is suppressed by the Klein-Nishina effect. If produced by electrons, the $\gamma$-ray emission is thus not expected to be peaked toward the stellar clusters, but rather trace the diffuse dust emission across the entire Cocoon. This adds difficulty to the task of distinguishing the leptonic and hadronic origins of the $\gamma$-ray radiation. The measurements of the Cygnus Cocoon emission above 10~TeV break the degeneracy of the two origins. As shown in the Extended Data Fig. 1, we find it unlikely that a single electron population produces $\gamma$ rays from GeV to the highest energy by inverse-Compton emission without its synchrotron radiation violating the flux constraints posed by  radio \cite{2003AJ....125.3145T} and X-ray \cite{suzaku} observations. The leptonic origin of the $\gamma$-ray radiation by the Cygnus Cocoon is thus disfavored, as uniquely responsible for the GeV and TeV flux observed.

The CR energy density above a proton energy of 10 TeV is calculated for four annuli up to 55~parsec from Cyg OB2 and is presented in the right panel of Fig.~2. We find that the CR energy density in all spatial bins is larger than the local CR energy density of $10^{-3} \rm \ eV/cm^3$ based on the AMS measurements \cite{localCR_density}. Therefore, as for the GeV $\gamma$ rays \cite{fermicocoon}, TeV $\gamma$ rays  also come from the freshly accelerated cosmic rays (CRs) inside the Cygnus Cocoon, rather than from the older Galactic population.

\begin{figure}
    \includegraphics[width=0.99\textwidth]{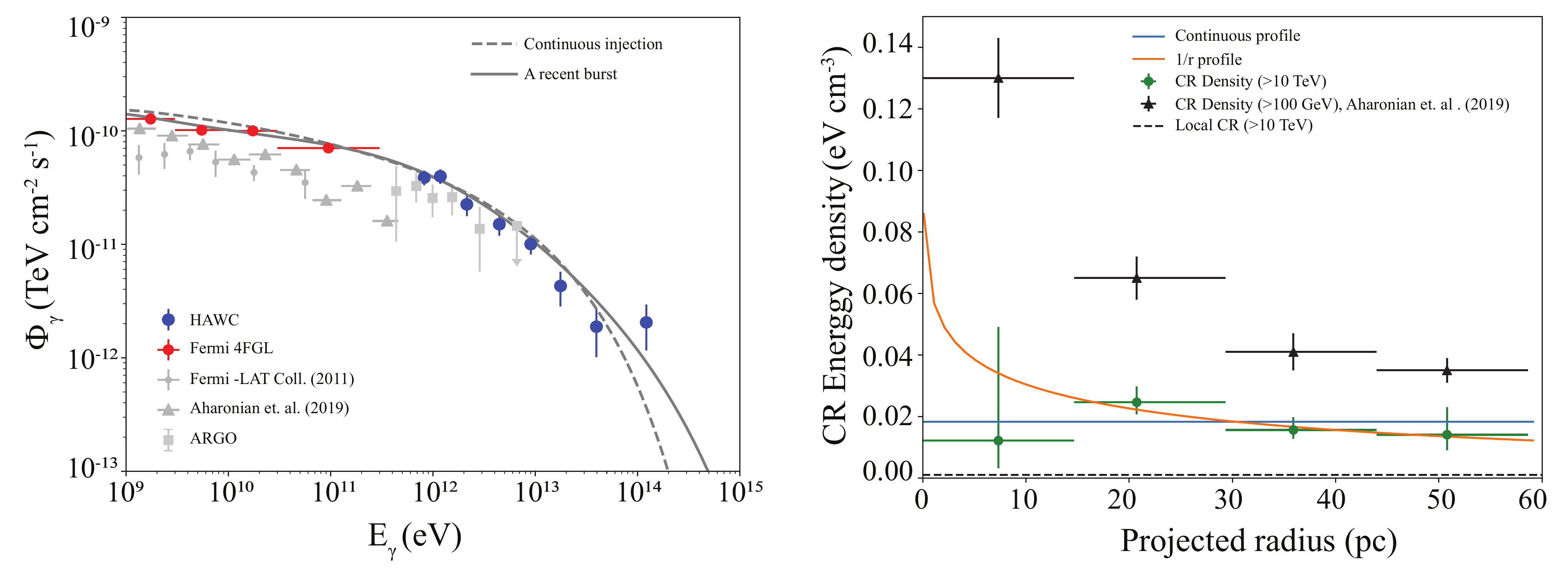}
    \caption{\textbf{Spectral energy distribution of the gamma-ray emission and cosmic ray (CR) density at the Cocoon region.} \textbf{Left}: Spectral energy distribution of the Cocoon measured by different $\gamma$-ray instruments. Here, $\Phi_{\gamma}$ is the $\gamma$-ray flux which is given by $E_{\gamma}^2 \times  \dfrac{dN}{dE_{\gamma}}$ and $E_{\gamma}$ is $\gamma$-ray energy. Blue circles  are the spectral measurements for the Cocoon in this study. The errors on the flux points are the 1$\sigma$ statistical errors. At low TeV energy, HAWC data agree with the measurements by the ARGO observatory shown in grey squares \cite{Bartoli14}. The red and grey circles are the {\it Fermi}-LAT flux points published in \cite{4fgl} and \cite{fermicocoon} respectively. The grey triangles are the {\it Fermi}-LAT analysis from \cite{aha_nature}. The thick grey solid and dashed lines are $\gamma$-ray spectra derived from the hadronic modeling of the region. The leptonic modeling is provided in Extended Data Fig. 1.
 \textbf{Right}: CR density profile calculated for four rings (([0:15], [15:29], [29:44] and [44:55]) pc) centered at the OB2 association. The Green circles are the CR density derived above 10 TeV using HAWC $\gamma$-ray data. The Y errors are the statistical errors and the X error bars are width of the X bins. The orange and the blue lines are the 1/r (signature of the continuous particle injection) and constant (signature of the burst injection) profiles respectively, calculated by assuming a spherical symmetry for the $\gamma$-ray emission region and averaging the density profile over the line of sight within the emission region. The black dashed line is the local CR density above 10 TeV based on the AMS measurements \cite{localCR_density}. The black triangles are the CR density above 100 GeV from \cite{aha_nature}.}
\end{figure}

The radial profile of the CR density yields information on the mechanism accelerating particles in the Cygnus Cocoon. Assuming that a CR accelerator has been active in the centre of the region at radius $(r)$=0, roughly at the location of Cyg OB2, a $1/r$ dependence of the CR density would imply that the acceleration process has continuously injected  particles in the region for 1-7 Myrs. A continuous acceleration process, which cannot be guaranteed by a single supernova explosion event, could be produced by the combined and long-lasting effect of multiple powerful star winds. On the other hand, a constant radial profile would imply a recent ($<$ 0.1 Myr) burst-like injection of CRs, such as a supernova explosion event. While the measured CR profile seems to agree with a $1/r$ dependence, a constant profile, namely a burst-like injection, cannot be excluded. This is in contrast to CR density profile above 100 GeV from \cite{aha_nature}, which clearly favors the $1/r$ profile. Alternatively, the $1/r$ profile is less striking for TeV CRs because of their escape time.

The angular size of the Cygnus Cocoon is about 2.1\degree, which translates into a radius r = 55 parsec at 1.4 kpc. The size of the Cocoon is similar in both the TeV and GeV energy range. Assuming a loss free regime, the particles from tens of GeV to hundreds of TeV diffuse in the region in a time $t_{\rm diff}$ given by $t_{\rm diff} = r^2/(2D)$ \cite{aharonian_book2}, where $D$ is the particle diffusion coefficient. If $D(E^*) = \beta\,D_0(E^*)$, where $D_0(E^*)$ is the average diffusion coefficient in the Galaxy at a given energy $E^*$ and $\beta$ is the suppression coefficient, then at 10 GeV 

$t_{\rm diff} (10 \ \rm GeV) = 15000 \times \left(\dfrac{1}{\beta}\right) \times \left(\dfrac{R_{\rm diff}}{55 \  \rm pc}\right)^2  \times \left(\dfrac{D_0(10 \ \rm GeV)}{3 \times 10^{28} \ \rm cm^2/s}\right)^{-1}$ yr.

The diffusion time ($t_{\rm diff}$) of 10 GeV particles detected with {\it Fermi}-LAT needs to be shorter than the age of the Cyg OB2 association $t_{\rm age}$, $t_{\rm diff} \,(10 \,\rm GeV) < t_{\rm age}\sim 1-7$~Myr \cite{agenew}, yielding $\beta > 0.002$. On the other hand, the diffusion time of 100 TeV particles must be longer than the light-travel time till the edges of the Cocoon, $t_{\rm diff} \,(100 \,\rm TeV) \gg R_{\rm diff} / c$, where \textit{c} is the speed of light. With $D_0(100 \rm {TeV}) = 3 \cdot 10^{30}$ cm$^2$/s, we obtain $\beta\ll 1$. The combination of observations by the GeV and TeV instruments provides unique insights to particle transport in the Cocoon super-bubble. The "suppression of the diffusion coefficient" ($\beta$) is found to be $0.002 < \beta \ll 1$. This confirms that closer to particle injectors, high turbulence is driven by the accelerated particles, and CRs are likely to diffuse slower than in other regions of the Galaxy.

As discussed in \cite{fermicocoon}, while the PWN powered by PSR J2021+4026 and PSR J2032+4127 cannot explain this extended Cocoon emission, we cannot rule out that the emission could be from a yet-undiscovered PWN. The nearby $\gamma$ Cygni Supernova remnant (SNR) might not have been able to diffuse over the Cocoon region because of its young age \cite{fermicocoon}. The $\gamma$-ray emission measured from the Cocoon region over five orders of magnitude in energy is likely produced by protons between GeV and PeV range colliding with the ambient dense gas. The spectral shape in the TeV energy range is well described by a power law without an indication of a cutoff up to energies above 100 TeV. Therefore, it might be the case that the powerful shocks produced by multiple strong star winds in the Cygnus Cocoon can accelerate particles, not only up to tens of TeV energies, as previously indicated by the {\it Fermi}-LAT detection, but even beyond PeV energies. However, the presence of a cutoff or a break in the GeV to TeV $\gamma$-ray spectrum at a few TeV, evidenced in the measurements of both ARGO and HAWC detectors, argues against the efficiency of the acceleration process beyond several hundred TeV.

 The break in the $\gamma$-ray spectrum around a few TeV could be either due to leakage of CRs from the Cocoon, or a cutoff in the CR spectrum injected from the source. In the first scenario, the $\gamma$-ray emission is dominated by recent starburst activities less than 0.1~Myr ago. The diffusion length in the Cocoon is 100-1000 times shorter than that in the interstellar medium due to strong magnetic turbulence \cite{fermicocoon}, plausibly driven by starburst activities. The lower-energy CRs are confined by the magnetic field of the Cocoon while higher-energy CRs escape from the region before producing $\gamma$ rays, resulting in a spectral break from GeV to TeV regime. An injection index, $\alpha\sim - 2.1$ for the CR spectrum is needed to explain the {\it Fermi}-LAT observation. Such a spectrum can be achieved by different particle acceleration mechanisms, for example through shock acceleration. An example of the leakage model is illustrated as the thick solid grey line in Fig. 2. Assuming a {recent activity that happens $0.1$~Myr ago}, 
 and a gas density of $30\,\rm cm^{-3}$ suggested by HI and HII observations \cite{butt}, the proton injection luminosity is found to  be $L_p \sim 4\times 10^{37}\,\rm erg\,s^{-1}$ above 1 GeV (see Methods).  The data above 100~TeV suggest that the stellar winds inject protons to above PeV with a hard  spectrum. 

In the second scenario, the $\gamma$-ray emission is produced by continuous starburst activities over the OB2 star lifetime, 1-7~Myr. 
In this scenario, a hard CR spectrum $\alpha\sim - 2.0$, depending on the index of the turbulence, is required to meet the $\gamma$-ray spectrum $\Gamma\sim - 2.1$ at GeV. As illustrated by the dashed grey curve in Fig.~2, a cutoff in the injected proton energy around 300~TeV can explain the change of spectral index from GeV to TeV regime. This scenario requires proton injection luminosity $L_p \sim 7\times10^{36}\,\rm erg\,s^{-1}$ above 1 GeV. 

The total mass of the OB2 association is $(2 - 10) \times 10^4 \rm M_\odot\ $ \cite{OB2mass, OB2_2010}, and the wind mechanical luminosity is estimated to be $\sim(1-2) \times 10^{39} \rm \, erg\,s^{-1}$ \cite{OB2mass}. The stellar association thus requires 4\% and 0.7\% acceleration efficiency for the burst and the steady model, respectively. OB2 produces sufficient power to account for the acceleration of CRs that now make up the Cocoon. 

The HAWC observation reveals the high-energy spectrum of Cygnus Cocoon, a representative of one of the most plausible Galactic CR source classes, at the highest $\gamma$-ray energies. The TeV measurements provide direct evidence that the Cygnus Cocoon accelerates CR protons above 100~TeV. In comparison to the GeV Cocoon \cite{fermicocoon},we don't observe a hard $\gamma$-ray spectrum above 1~TeV. While the $\gamma$-ray emission of superbubbles turns out to be more complicated than previously understood, we show that the Cocoon may still be a PeVatron if the break in the energy spectrum is caused by the escape of higher-energy CRs. Our result suggests that Cygnus Cocoon could also emit high-energy neutrinos created by the decay of ions produced by hadronic interactions of the CR protons with the gas density in the ambient region. While these neutrinos have not been detected yet, they may be identified through extended source analysis by IceCube and future neutrino experiments \cite{icecube_ex, icecube_gen2}. The feasibility of these neutrino detections could be evaluated using a hadronic $\gamma$-ray emission template based on the TeV Cocoon model. Future VHE $\gamma$-ray observations such as SWGO \cite{swgo} and LHAASO \cite{lhaaso} will provide more statistics that can be used to resolve the contribution of other stellar clusters to Galactic CRs around the knee.

{\textbf{Acknowledgements} We acknowledge the support from: the US National Science Foundation (NSF); the US Department of Energy Office of High-Energy Physics; the Laboratory Directed Research and Development (LDRD) program of Los Alamos National Laboratory; Consejo Nacional de Ciencia y Tecnolog\'ia (CONACyT), M\'exico, grants 271051, 232656, 260378, 179588, 254964, 258865, 243290, 132197, A1-S-46288, A1-S-22784, c\'atedras 873, 1563, 341, 323, Red HAWC, M\'exico; DGAPA-UNAM grants IG101320, IN111315, IN111716-3, IN111419, IA102019, IN112218; VIEP-BUAP; PIFI 2012, 2013, PROFOCIE 2014, 2015; the University of Wisconsin Alumni Research Foundation; the Institute of Geophysics, Planetary Physics, and Signatures at Los Alamos National Laboratory; Polish Science Centre grant, DEC-2017/27/B/ST9/02272; Coordinaci\'on de la Investigaci\'on Cient\'ifica de la Universidad Michoacana; Coordinaci\'on General Acad\'emica y de Innovación (CGAI-UDG; SEP-PRODEP-UDG-CA-499) Royal Society - Newton Advanced Fellowship 180385; Generalitat Valenciana, grant CIDEGENT$/2018/034$. Thanks to Scott Delay, Luciano D\'iaz and Eduardo Murrieta for technical support.

We thank Seth Digel for helpful discussion regarding the source modeling of the Cygnus Cocoon region in the {\it Fermi} 4FGL catalog.
}

\textbf{Authors Contribution:}
B. Hona (bhona@mtu.edu) analyzed the HAWC data, performed the maximum-likelihood fit of the multi-source model, hadronic model fit and CR density study. H. Fleischhack (hfleisch@mtu.edu) and P. Huentemeyer (petra@mtu.edu) helped in the development of the multi-source model and in the scientific interpretations of the fit results. K. Fang (kefang@stanford.edu) and R. Blandford (rdb3@stanford.edu) helped  develop the hadronic emission models and helped in the interpretations of the model fit results. K. Fang also developed the leptonic emission model and provided its interpretations. S. Casanova (sabrinacasanova@gmail.com) motivated the Cocoon analysis, helped with the interpretations of the leptonic model and performed the diffusion coefficient suppression study at the Cocoon region. B. Hona, K. Fang and S. Casanova prepared the manuscript. The full HAWC Collaboration has contributed through the construction, calibration, and operation of the detector, the development and maintenance of reconstruction and analysis software, and vetting of the analysis presented in this manuscript. All authors have reviewed, discussed, and commented on the results and the manuscript.

\clearpage

\section{Extended Data Figure}

\renewcommand{\figurename}{Extended Data Figure}
\renewcommand{\thefigure}{\arabic{figure}}

\setcounter{figure}{0}
\begin{figure}
\includegraphics[scale=0.5]{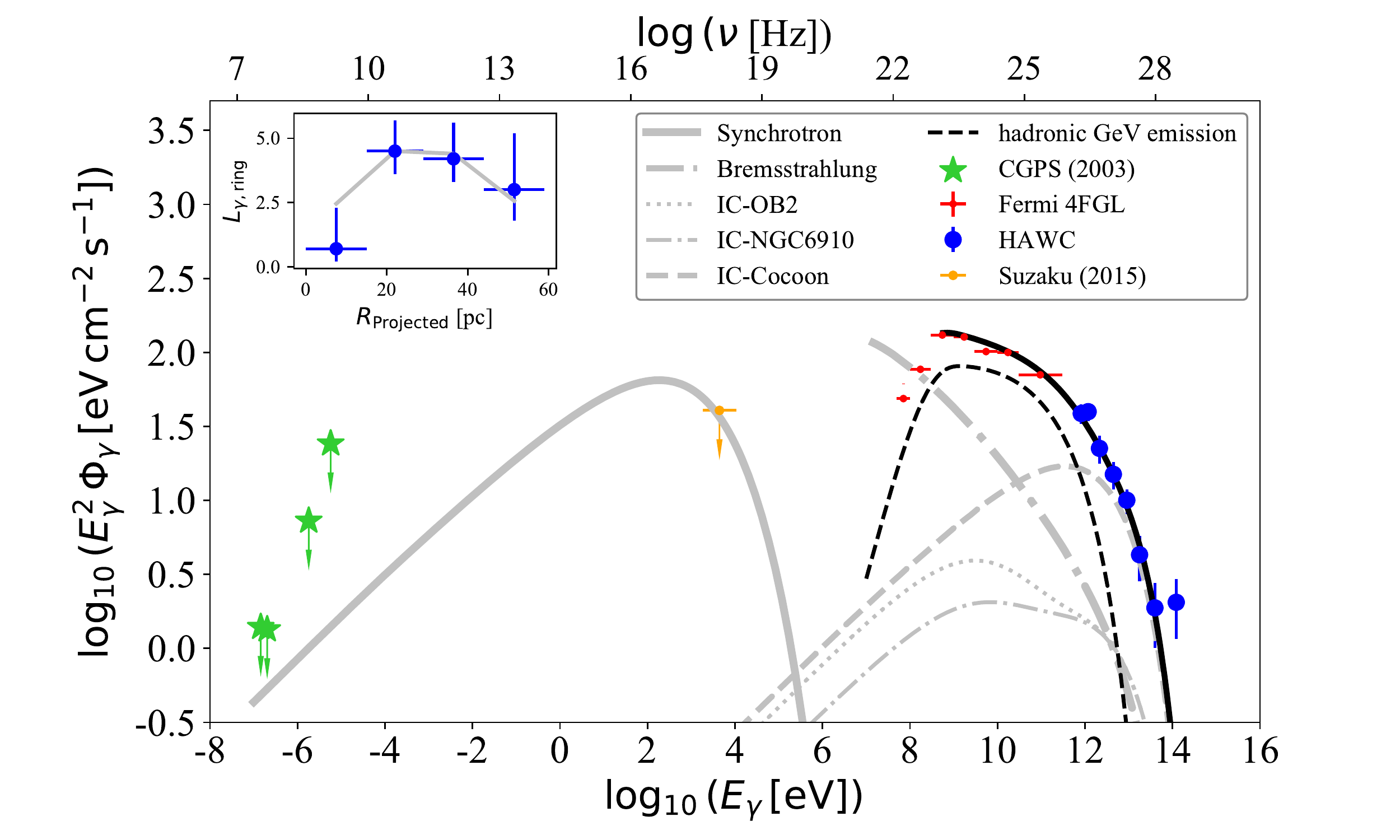}
\captionsetup{width=.9\linewidth}
\caption{ \textbf{Leptonic modelling at the Cocoon region.} Multi-wavelength observations of the Cygnus Cocoon \cite{4fgl, suzaku, aha_nature} constrain the Synchrotron and Bremsstrahlung radiation of relativistic electrons. The light grey curves correspond to a ``minimum leptonic model", where only $\gamma$-rays above 1~TeV are explained by electron emission. The electron population is assumed to follow a power-law energy spectrum $dN/dE\propto E^{-2}$ in a region with magnetic field $B=20\,\mu$G and gas density $n=30\,\rm cm^{-3}$ as in the Cocoon \cite{fermicocoon}. The leptonic emission consists of the Synchroton radiation (solid, from radio to hard X-ray), Bremsstrahlung emission (thick dash-dotted), and inverse-Compton scattering of the dust emission in the Cocoon (dashed) and the radiation fields of the two stellar clusters, NGC~6910 (thin dash-dotted) and OB2 (dotted). Observations between 0.1--100~GeV are explained by hadronic interaction (black dashed curve). The red points are the GeV flux points by {\it Fermi}-LAT and the blue circles are the HAWC flux points with 1 $\sigma$ statistical errors. The sum of the emission above $\sim$0.3~GeV is indicated by the black solid curve.  In the inner plot,  the blue circles indicate the $\gamma$-ray luminosity for the four rings at the Cocoon region and the light grey solid curve is the TeV $\gamma$-ray luminosity from the model.}
\label{fig:leptonic}
\end{figure}

\clearpage

\section{METHODS}

\subsection{The HAWC Observatory}
The HAWC $\gamma$-ray instrument is comprised of an array of 300 water Cherenkov detectors (WCDs) at an altitude of 4100 m in Sierra Negra, Mexico. It is sensitive to $\gamma$ rays in the energy range of a few hundreds GeV to beyond 100 TeV \cite{hawccrab_new1}. Each WCD has four photomultiplier tubes (PMT) at the bottom which detect the Cherenkov light produced by air shower particles travelling through the WCD. Air shower events recorded by the detector are reconstructed to extract shower properties. The hadronic CRs that pass gamma/hadron separation cuts during the reconstruction are the main background in the analysis of $\gamma$-ray sources \cite{Abey17crab1}. The background is computed in the same manner as \cite{hawccrab_new1}.  \

HAWC data are divided into 9 size bins according to the fraction of PMTs triggered in a shower event. Size bin 9 has the highest-energy $\gamma$ rays and the best angular resolution of 0.17\degree{} (68\% containment radius) or better \cite{Abey17crab1}. Each size bin is further subdivided into 12 quarter decade energy bins for a total of 108 bins \cite{hawccrab_new1} using the "ground parameter" energy estimation algorithm, which uses charge density at 40 m from the shower axis \cite{hawccrab_new1}. 
The results presented here are using 1343 days of data collected between Jun. 2015 to Feb. 2019 for events with energies that were reconstructed above 1 TeV by the "ground parameter" method \cite{hawccrab_new1}.

\subsection{Model fitting at the Cocoon Region}
A model describing the sources in the Cocoon region is fitted using the maximum likelihood code of 3ML \cite{3ml1}. This method estimates the best fit value for the parameters being fitted for the spatial and spectral description of the model in order to maximise the likelihood of the model. Given the model, the expected $\gamma$-ray events and background events are calculated by forward folding with the detector response. A test statistic (TS) of each source is then defined as the log of ratio of maximum likelihood of the best-fit model including the source $L_{(\rm source)}$ in question to the likelihood of the best-fit model without that source $L_{(\rm no-source)}$ :

\begin{equation}
\label{eq:TS}
TS = 2 \ln \dfrac{L_{(\rm source)}}{L_{(\rm no-source)}} .
\end{equation}

The region of interest used for the fit is a disk of radius 6\degree{} centered at (RA = 307.17, Dec = 41.17)\degree{} with 2\degree{} mask around the bright Cygnus region source 2HWC J2019+367 \cite{catalog} (eHWC J2019 + 368 \cite{HE_catalog1}). The best fit description of the $\gamma$-ray emission in the ROI includes three sources. Two sources, HAWC J2031+415 and HAWC J2030+409 together contribute to the emission detected at 2HWC J2031+415 region \cite{Hona:2019ysf1}. HAWC J2031+415 is possibly associated with the PWN TeV J2032+4130, which is the first extended $\gamma$-ray source detected in the VHE range \cite{pwn2002}. After the discovery by HEGRA, the detection has been confirmed by various observatories \cite{whipple1, MAGIC1, pwnmilagro1, veritas2014}. The $\gamma$-ray emission is likely a PWN, powered by PSR J2032+4127 \cite{veritas2014}. HAWC J2030+409 is a TeV counterpart of the {\it Fermi}-LAT Cocoon \cite{fermicocoon}. The ROI also includes a nearby third source 2HWC J2020+403 ($\gamma$ Cygni), which is possibly associated with VER J2019+407 \cite{verj20191}, enclosed within the radio shell of the $\gamma$ Cygni SNR. The three sources account for all the TeV emission observed in the ROI region (see Supplementary Fig. 2). 
The spectral fit results for these three sources along with the TS values are provided in Extended Data Fig. 2. The profile of the emission centered at the HAWC J2030+409 is given in Supplementary Fig. 1b, where the blue line is the model after subtracting PWN and gamma Cygni.

Different spectral models were explored for HAWC J2030+409 (HAWC Cocoon): a power law  spectrum,
\begin{equation}
\label{eq:pl}
\dfrac{dN}{dE}=N_{0}\left(\dfrac{E}{ E_{0}}\right)^{\Gamma},
\end{equation}
and a power law with an exponential cutoff,
\begin{equation}
\label{eq:plc}
\dfrac{dN}{dE}=N_{0}\left(\dfrac{E}{E_{0}}\right)^{\Gamma} \times \ \exp{\left(-E/E_{c}\right)}.
\end{equation}
\
The pivot energy $E_{0}$ is fixed at 4.2 TeV for both equations (\ref{eq:pl}) and (\ref{eq:plc}). The free parameters are $N_{0}$ and $\Gamma$. Additionally, the Gaussian width is also fitted. The two equations (\ref{eq:pl}) and (\ref{eq:plc}) constitute nested models. Using Wilks' theorem \cite{wilks1} to convert $\Delta$TS to a significance, there is no significant improvement of TS value for a spectral fit using Eq.~\ref{eq:plc} (TS = 196.9) compared to Eq. \ref{eq:pl} (TS = 195.2). Hence, for the HAWC J2030+409 spectrum, there is no significant preference of cutoff in the spectrum in comparison to a simple power law spectrum. The TeV Cocoon spectrum is thus described by a power law spectrum. The likelihood profile of the cutoff energy assuming a power law spectrum with an exponential cutoff for the Cocoon is shown in Supplementary Fig. 1a. Using 95\% confidence level, the lower limit to a cutoff is obtained at 15 TeV.

The HAWC J2031+415 (PWN) is described by a power law spectrum with an exponential cutoff and Gaussian morphology. A power law spectrum with an exponential cutoff is chosen since it is preferred in our model in comparison to a power law spectrum by $\Delta \rm TS \rm \ of \  16.9 \ (\sim 4 \sigma)$. The TS values for the PWN using a simple power law spectrum and a power law with an exponential cutoff are 281.6 and 298.5 respectively. The free parameters are flux normalization $N_{0}$, index $\Gamma$ and cutoff energy $E_{c}$ of Eq. (\ref{eq:plc}) plus the Gaussian width. The 2HWC J2020+403 ($\gamma$ Cygni) is described by a power law spectrum with a disk radius fixed at 0.63\degree {} based on the studies of \cite{henrike1}. The free parameters are $N_{0}$ and $\Gamma$ of Eq. (\ref{eq:pl}). \\

\subsection{Residual Significance Distribution} 
Supplementary Fig. 2 shows the significance distribution in the ROI before and after subtracting various sources. If the residual map obtained after subtracting the model contains only background fluctuations, then the significance histogram should follow a normal distribution. The dotted lines in the plots are the expected and the obtained distribution.

Shown in Fig. 1 is the significance map with the ROI circled in green. The significance distribution obtained for the ROI as shown in Supplementary Fig. 2a is skewed towards positive values due to the presence of different $\gamma$-ray sources. Because of these sources, we see excess counts above background fluctuations. After subtraction of the PWN and the $\gamma$ Cygni SNR, Supplementary Fig. 2b  shows considerably reduced skew. The excess counts from the Cocoon source contribute to this skew.  Finally, in Supplementary Fig. 2c in which the Cocoon source in addition to the PWN and $\gamma$ Cygni are subtracted, there is no longer significant excess counts over background fluctuations. 

\subsection{Energy Range}
A study similar to the energy range method in \cite{geminga1} is performed to determine the energy range of $\gamma$-ray emission from the TeV Cocoon. The best fit model (the power law spectrum, Equation \ref{eq:pl}) is multiplied by a step function at some value to simulate a sharp cutoff in energy. The free parameters are the strict upper or lower cutoff for the energy in addition to the flux normalisation and index of the Cocoon. The energy value where the log likelihood decreases by 1$\sigma$ from the maximum log likelihood value in the nominal case (see Supplementary Fig. 3) is then quoted as the lower limit to the maximum detected $\gamma$-ray energy (when the free parameter is a strict upper cutoff) or the upper limit to the minimum detected $\gamma$-ray energy (when the free parameter is a strict lower cutoff). 

Based on the energy range studies, the TeV Cocoon spectrum extends from 0.75 TeV to 225 TeV. The TS values for the Cocoon in each median reconstructed energy are provided in Supplementary Table 1. The TS above 100 TeV is about 6.

\subsection{CR density profile}
 The annular rings used in the density profile study are similar to the annular bins used in \cite{aha_nature}. The four rings [0:0.6]\degree, [0.6:1.2]\degree, [1.2:1.8]\degree{} and [1.8:2.2]\degree{} (corresponding to [0:15] pc, [15:29] pc, [29:44] pc and [44:55] pc) centered at the position of OB2 association (308.3, 41.3)\degree {} are selected as illustrated in Fig. 1. Using the 3ML software \cite{3ml1}, the four rings are fit simultaneously with the contributions from PWN and $\gamma$ Cygni. In total, this six-source (four rings, PWN and $\gamma$ Cygni) model has 13 free parameters (flux normalisation and index of the four rings, flux normalisation, index and cutoff energy of the PWN plus flux normalisation and index of $\gamma$ Cygni). Using the integral flux $(I_{ring})$ for each ring from $\sim$ (1 TeV to 200 TeV), the total luminosity of each ring is calculated as
\begin{equation}
L_{\gamma}= 4\pi I_{ring} \times d^2,
\label{eq:luminosity}
\end{equation}
where  $d$ is the distance to the OB2 association (1.4 kpc \cite{ob2distance1}). Gas mass ($M$) in the region is also used as quoted in \cite{aha_nature} and the CR density can be calculated using the formula in Eq. \ref{eq:density} \cite{2016Natur.531..476H}.

\begin{equation}
w_{CR} (>\rm 10TeV) = 1.8\times10^{-2}\left(\dfrac{\eta}{1.5}\right)^{-1} \dfrac{L_{\gamma} (\geq 1TeV)}{10^{34} \rm erg\,s^{-1}} \left(\dfrac{M}{10^{6} M_{\odot}}\right)^{-1} \rm eV/cm^3
\label{eq:density}
\end{equation}

 $\eta$ accounts for presence of the heavier than hydrogen nuclei and is used as 1.5 \cite{for_eta11, for_eta21}. Then, using Eq. (\ref{eq:density}), the density values $w_{CR}$ above 10 TeV are given in Supplementary Table 2. Shown in Fig. 2, the green circles represent CR energy density against the distance from the center of the OB2 association. The average CR density profiles over the line of sight are shown as orange and blue lines in Fig. 2. The reduced $\chi^2$ is 1.12 for a constant profile and 0.46 for 1/r profile. Due to the large statistic errors, the study cannot provide conclusive evidence of 1/r signature for continuous injection vs constant profile for burst-like injection. According to \cite{aha_nature}, the systematic uncertainties associated with the gas mass in the Cocoon region could be as high as 50\%. Adding the systematic error of +/-50$\%$ in the gas mass does not alter our conclusion that our study cannot claim a preference for continuous injection vs burst-like injection.

\section{Hadronic modeling}

Protons interact with the ambient gas cloud and produce $\pi^{0}$ which immediately decays into $\gamma$ rays. Fig. 2 {} shows the expected $\gamma$-ray flux from the parent proton flux in the Cocoon. We assume 59.2 mb cross section for the pp interaction between a PeV proton and a rest-mass proton \cite{kelner1}. The gas density of the Cocoon region is approximately 30 nucleons/$\rm cm^3$ \cite{butt}.

\subsection{Transient Source}
Here, we consider a starburst event at $t_0$ that injects particles with rate $Q(E, t) = S(E)\,\delta(t-t_0)$, where $S(E) \equiv S_0 E^{-\alpha}\exp{(-E/E_{\rm cut})}$ depends only on energy and $\alpha$ is spectral index. The solution of the CR transport equation applied to the Cocoon region is \cite{burstsolution1} 
\begin{eqnarray}
    n(E, r, t - t_0)
   = \frac{b(E_0)}{b(E)} S(E_0)\,\frac{\exp{[-r^2 / (4\lambda_0^2)}]}{\left(4\pi\lambda_0^2\right)^{3/2}},
   \label{eqn:np_t}
\end{eqnarray}
where $E_0$ is the initial proton energy at $t_0$ and $b(E)= dE/dt$ is the energy loss rate. $\lambda_0$ is the distance traveled by the particle when its energy decreases from $E_0$ to $E$ (or, from time $t_0$ to time $t$) and is a function of diffusion coefficient $D(E)$,
\begin{equation}
    \lambda_0 = \left[\int_E^{E_0} \,d\epsilon\, \frac{D(\epsilon)}{b(\epsilon)}\right]^{1/2} \ .
\end{equation}

Assuming that the proton density inside the Cocoon is from a source at the center, the total number of CRs from a burst at $t_0$ is 
\begin{eqnarray}
    \frac{dN}{dE}(t- t_0)&=&\frac{b(E_0)}{b(E)}S(E_0) \,F_{\rm cc}(E, t_0)\\
    &=&\dfrac{E_0}{E}S_0E^{-\alpha}\exp\left(-E/E_{cut}\right)\,F_{\rm cc}(E, t_0) 
\end{eqnarray}
where $F_{\rm cc}$ depends on the radius of the Cocoon $(R_{\rm cc}$ = 50 pc) and the time of the burst,
\begin{eqnarray}
    F_{\rm cc}= {\rm erf}\left(\frac{R_{\rm cc}}{2\lambda_0}\right) - \frac{1}{\pi^{1/2}}\frac{R_{\rm cc}}{\lambda_0}\exp{[-R_{\rm cc}^2/(4\lambda_0^2)]}\ .
\end{eqnarray}

For the turbulent magnetic field in the Cocoon, $D(E) = 10^{25} (E/1 \rm GeV)^{0.55} {} \ \rm cm^2 \ s^{-1}$ is adopted. Fitting the function with the 4FGL Cocoon flux points \cite{4fgl} and HAWC data, the initial injection rate $S_{0}$ for $t_{0}$ = 0.1 Myr obtained is  $\sim$ $9 \times 10^{37}$ $\rm erg\,s^{-1}$. We find the best fitted injected proton spectrum with spectral index  $\alpha$ = $-$2.13 with $E_{\rm cut}$ fixed at $\sim$ 100 PeV.  The result is insensitive to the value of $E_{\rm cut}$ as long as it is above $\sim 1$~PeV.

The proton luminosity ($L_p$) is then calculated as 
\begin{eqnarray}
  L_p = \int_{E_{min}}^{E_{max}} {Q_0E^{-\alpha} \rm exp (-E/E_{cut}) dE} \ .
  \label{eq:p_luminosity}
\end{eqnarray}

For $E_{min}$ = 1 GeV and $E_{max}$ = 1 PeV, $L_p \sim 4\times 10^{37}\,\rm erg\,s^{-1}$.

\subsection{Steady Source}

If the source injects particles continuously in time with a temporal profile $Q(t)$, the CR flux sums the contribution from different injection episodes.  
\begin{equation}
    \frac{dN}{dE}(t) = \int_0^{\tau_{\rm cc}} dt_0\,\frac{dN}{dE}(t-t_0)Q(t_0) 
\end{equation}
where $\tau_{\rm cc}$ is the source age. 
Using $\tau_{\rm cc}$ = 3 Myr, the initial injection rate $Q(t_{0})$ obtained is $\sim 1 \times 10^{36}$ $\rm erg\,s^{-1}$. The best spectral fit has spectral index $\alpha$ = $-$2.0 and cutoff $E_{\rm cut}$ fixed at $\sim$ 300 TeV. For the steady source, $D(E) = 10^{25} (E/1\rm GeV)^{0.33} {} \ \rm cm^2 \, s^{-1}$ is adopted. Using $E_{min}$ = 1 GeV and $E_{max}$ = 300 TeV, $L_p  \sim 7 \times 10^{36}\,\rm erg\,s^{-1}$. 

Assuming $(0.7 -4)\%$ acceleration efficiency from the proton injection luminosity obtained here for the two models, to generate $ (0.3- 1) \times 10^{41} \,\rm erg\,s^{-1}$  power to explain the energy density of Galactic CRs  \cite{drury1}, $\sim$ $10^4$ Cygnus OB2-like stellar associations would be required. Currently, $<$ 100 OB associations (Cygnus OB2 being the most massive) have been identified in the 3 kpc survey of our Galaxy \cite{oblist11, oblist21}.

Both of these models (Transient and Steady source) adequately describe the observed GeV to TeV data. Given the current statistical errors, we can not provide conclusive evidence of one model preference over the other.

The $\gamma$-ray flux produced by protons injected at $t_0$ is
\begin{equation}
    \Phi_\gamma (E_\gamma,t) = \frac{c\,n_H}{4\,\pi\, D_{\rm cc}^2}\int_{E_\gamma}^{\infty} dE_p \sigma_{\rm pp}(E_p) \left[F_\gamma(\frac{E_p}{E_\gamma}, E_p)\,\frac{1}{E_p}\right]  \frac{dN_p}{dE_p}(t-t_0), 
\end{equation}
where $F_\gamma(E_p/E_\gamma, E_p)$ is the spectrum produced by one proton with energy $E_p$ via $\pi^0$ decay. An analytical form of $F_\gamma$ is presented in equation 58 of \cite{kelner1}. 

The transition of propagation regimes \cite{2015PhRvD..92h3003P1} may also lead features in a $\gamma$-ray spectrum, but cannot explain the  spectral change around 1~TeV observed from the Cygnus Cocoon since proton propagation at these energies is still in the diffusive regime.

\section{Leptonic Modeling}

The electron spectrum is computed by solving the following transport equation 
\begin{equation}
    \frac{\partial n(E, r)}{\partial t} - \nabla [D(E)\,\nabla n(E, r)] - \frac {\partial}{\partial E}[b(E)\,n(E, r)] = Q(E, t)\,\delta(r)
\end{equation}
where $D(E)=D_0\,E^\delta$ is the diffusion coefficient with $\delta\approx 0.33$ for the Kolmogorov turbulence, $Q(E, t)=Q_0\,E^{-\alpha}$ is the particle injection rate, and $b(E)= dE/dt$ is the energy losing rate. Assuming that the turbulent magnetic field and gas density inside the Cocoon are roughly constant over time, $D$ and $b$ depend only on energy. The computation of the energy losing rate is described below. For selected parameter values of $D_0$, $Q_0$ and $\alpha$, the gamma-ray produced by the electrons are set to explain both the spatial profile and the energy spectrum measured by HAWC.

Following \cite{fermicocoon}, we consider three radiation fields for $\gamma$-ray production, including intense stellar light fields surrounding Cyg~OB2 and NGC~6910 (a stellar cluster in the vicinity of OB2), and a more diffuse dust radiation field that spans the entire Cocoon. We note that the stellar radiation fields are not important to TeV $\gamma$-ray production. The energy of a stellar photon in an electron's rest frame is $\epsilon' / (m_e\,c^2)\sim 6\,(E_e/1\,\rm TeV)\,(T/2\times10^4\,\rm K)\gg 1$, where $E_e$ is electron energy and $T$ is temperature of the stars. The inverse-Compton emission at TeV is thus largely suppressed by the Klein-Nishina effect. The dust radiation field on the other hand peaks at lower frequency and can be upscattered to TeV $\gamma$-rays. The radiation field is also more extended than the main clusters of Cyg~OB2 and NGC~6910. The extended distribution of $\gamma$ rays alone cannot reject a leptonic scenario where $\gamma$ rays are produced by electron-positron pairs. 

We compare the synchrotron, Bremsstrahlung, and inverse-Compton emission of high-energy electrons with multi-wavelength observations of the Cygnus Cocoon. We find it implausible that a single electron population explains both the GeV to TeV flux and the spatial profile of the $\gamma$ rays simultaneously, while not violating the X-ray and radio upper limits. Here, we presents a ``minimum leptonic model" (see Extended Data Fig.1) where only $\gamma$-ray emission above $\sim$1~TeV is explained by electrons. The observed flux between 0.1--100~GeV \cite{4fgl}   is assumed to be produced by hadrons such that they don't contribute to Synchrotron radiation at lower energy. The Extended Data Fig. 1 shows that even this minimum model can hardly satisfy the constraints posed by multi-wavelength observations. 
 
Deep X-ray observations of the $\gamma$-ray  Cocoon in the 2-10~keV range \cite{suzaku} and radio flux averaged over the Cocoon from the CGPS survey \cite{2003AJ....125.3145T} independently constrain the synchrotron emission by relativistic electrons.  In addition, relativistic electrons that may explain the HAWC observation between 1-100~TeV would unavoidably over produce a sub-GeV flux of Bremsstrahlung radiation.   We therefore conclude that the observed TeV $\gamma$ rays are more plausibly produced by hadrons. We focus on TeV $\gamma$-ray observation in this work, and refer to \cite{aha_nature, fermicocoon} for discussion about the origin of the observed GeV $\gamma$-rays. Future hard X-ray and $\gamma$-ray observations resolving subregions of the Cocoon will allow further investigation of the $\gamma$-ray production mechanism, which in principle could be a combination of leptonic and hadronic contributions from various stellar activities in the history of the Cocoon.

\section{Summary of systematic uncertainties}
The contribution to the systematic uncertainties from the detector effects to the flux normalisation of the TeV Cocoon is about $\pm 7\%$ of the nominal value. The index and Gaussian width of the TeV Cocoon change by $< 2\%$ due to detector systematics. These uncertainties are determined in manner described in \cite{hawccrab_new1}. Further systematic studies are done with a larger ROI without masking 2HWC J2019+367. Using a larger ROI, the TeV Cocoon flux normalisation and the Gaussian width differed by $< 4\%$. The index of the TeV Cocoon differs by $< 1\%$. To explore the possible contamination from Galactic diffuse emission + unresolved sources, in addition to the three sources mentioned, a large diffuse emission background is also included. It is included in the model either as a Gaussian (spatial morphology) distribution symmetrically distributed about Galactic latitude of 0\degree{} and infinitely extended along Galactic longitude or as a uniform background with disk radius of 6\degree. In both cases, the additional (Galactic diffuse emission + unresolved sources) component is not significantly detected. However, its presence could decrease the Cocoon flux by about 11\%. The effects on the index and the Gaussian width are negligible ($<$ 2\%). In the GeV regime, we studied the Cocoon spectrum published in a 2011 \cite{fermicocoon} paper, 3FHL \cite{3fhl1} and 4FGL \cite{4fgl} catalog. The 3FHL and 4FGL catalog report 10-15\% higher flux for the Cocoon compared to the published spectrum in \cite{fermicocoon}. The results reported in this study are based on the measurements from the 4FGL catalog \cite{4fgl}. Fitting the proton spectrum using the flux points from \cite{fermicocoon} and \cite{3fhl1} and HAWC data results in no significant difference for the total energy of the protons in the Cocoon. 

\textbf{Data Availability:} Currently, the datasets analyzed during this study and the scripts used are available at a public data repository (https://github.com/binitahona/Cocoon-paper).

\textbf{Code Availability:} The study was carried out using the Analysis and Event Reconstruction Integrated Environment Likelihood Fitting Framework (AERIE-LiFF), the Multi-Mission Maximum Likelihood (3ML)
software, and the HAWC Accelerated Likelihood (HAL) framework. The code is open-source and publicly available on Github: https://github.com/rjlauer/aerie-liff,  https://github.com/giacomov/threeML and https://github.com/threeML/hawc\_hal. The software includes instructions on installation and usage.
\linebreak

\textbf{Competing interests:}
The authors declare no competing interests.

\clearpage

\section{Supplementary Information}
\section{Supplementary Tables}

\renewcommand{\figurename}{Supplementary Figure}
\renewcommand{\thefigure}{\arabic{figure}}
\renewcommand{\tablename}{Supplementary Table}

\setcounter{figure}{0}
\hfill
\begin{table*}[ht]
\centering
\begin{tabular}{c c c c}
    \hline
    \textbf{Rings (pc)} & \textbf{Mass}  & $\mathbf{L_{\gamma}(>1 TeV) } $  & $\mathbf{w_{CR}(>10TeV) }$\\ 
    & ($10^5$ \(M_\odot\)) & ($10^{33} \ \rm erg \cdot \rm s^{-1}$) & $\rm eV/cm^3$\\
    \hline
    &&&\\
      $0<r<15$ & 0.8 & $0.5_{- 0.4}^{+ 1.6}$ & $0.012_{- 0.090}^{+ 0.037}$\\
      &&&\\
      $15<r<29$ & 2.4& $3.3_{-0.5} ^{+ 0.7}$ & $0.025_{ - 0.004} ^{ + 0.005}$\\
      &&&\\
      $29<r<44$ & 4.0& $3.5_{ - 0.6}^{ + 0.9}$  & $0.016_{ - 0.003} ^{+ 0.004}$\\
      &&&\\
      $44<r<55$  & 3.3 & $2.6_{- 0.9}^{ + 1.6}$ & $0.014_{- 0.004}^{+ 0.009}$\\
    \hline  
\end{tabular}
\captionsetup{width=.9\linewidth}
\caption{\textbf{Luminosity and CR Density values in the four rings at the Cocoon region}. The given uncertainties are 1 $\sigma$ statistical errors. The gas mass is quoted from (16).}
\label{tab:densityvalues}
\end{table*}

\hfill

\begin{table*}[ht]
\centering
\resizebox{0.99\linewidth}{!}{  
\begin{tabular}{c c c c c c c} 
    \hline
    &&&& \textbf{Spectral Parameters}\\
    \textbf{Source} & \textbf{Spectral Model} & \textbf{TS} &  $N_{0}$ & $E_{0}$ & $\Gamma$ &$E_{c}$   \\ 
    
    &&&  $({\rm TeV^{-1} cm^{-2} s^{-1}})$ &(TeV) & & (TeV) \\
    \hline 
     &&&&&&\\ 
      HAWC J2030+409 & Eq. 2 & 195.2  & $9.3_{-0.8}^{+0.9} \times 10^{-13}$ & 4.2 & $-2.64_{-0.05}^{-0.05}$ & \\ 
      
      &&&&&&\\
      HAWC J2031+415 & Eq. 3 & 298.5  & $1.3_{-0.2}^{+0.2} \times 10^{-13} $ & 4.9 & $-1.90_{-0.16}^{-0.20}$ & $33_{-10}^{+17}$  \\
      
      &&&&&& \\
      2HWC J2020+403 & Eq. 2 & 53.7  & $39_{-6}^{+6} \times 10^{-13} $ & 1.1 & $-2.95_{-0.13}^{-0.12}$ & \\
      
    \hline  
\end{tabular}
}
\captionsetup{width=.9\linewidth}
\caption{\textbf{Maximum likelihood fit results for the three sources in the ROI}. All uncertainties quoted are 1 $\sigma$ statistical errors.}
\label{tab:allfitvalues}
\end{table*}

\begin{table*}[ht]
\centering
\begin{tabular}{c c c}
    \hline
    \textbf{Energy Bins (TeV)} & \textbf{Median Energy (TeV)} & \textbf{TS}\\   
   \hline 
      1.0 - 1.78 & 0.83& 57.83\\
      1.78 - 3.16 & 1.20& 66.19\\
      3.16 - 5.62 & 2.18& 30.10\\
      5.62 - 10.0 & 4.50& 49.34\\
      10.0 - 17.8 & 9.08& 40.50\\
      17.8 - 31.6 & 17.73& 13.94\\
      31.6 - 100 & 39.79& 11.47\\
      100 - 316 & 122.18& 6.31\\
    \hline  
\end{tabular}
\captionsetup{width=.9\linewidth}
\caption{\textbf{Test statistics (TS) of the Cocoon in each energy bin}. The median energy quoted here is the HAWC reconstructed energy.}
\label{tab:energybins_TS}
\end{table*}

\hfill 
\hspace{3cm}
\newpage
\hfill

\clearpage

\section{Supplementary Figures}

\begin{figure} [hbt!]
\begin{subfigure}{.5\textwidth}
 \includegraphics[width=0.9\textwidth]{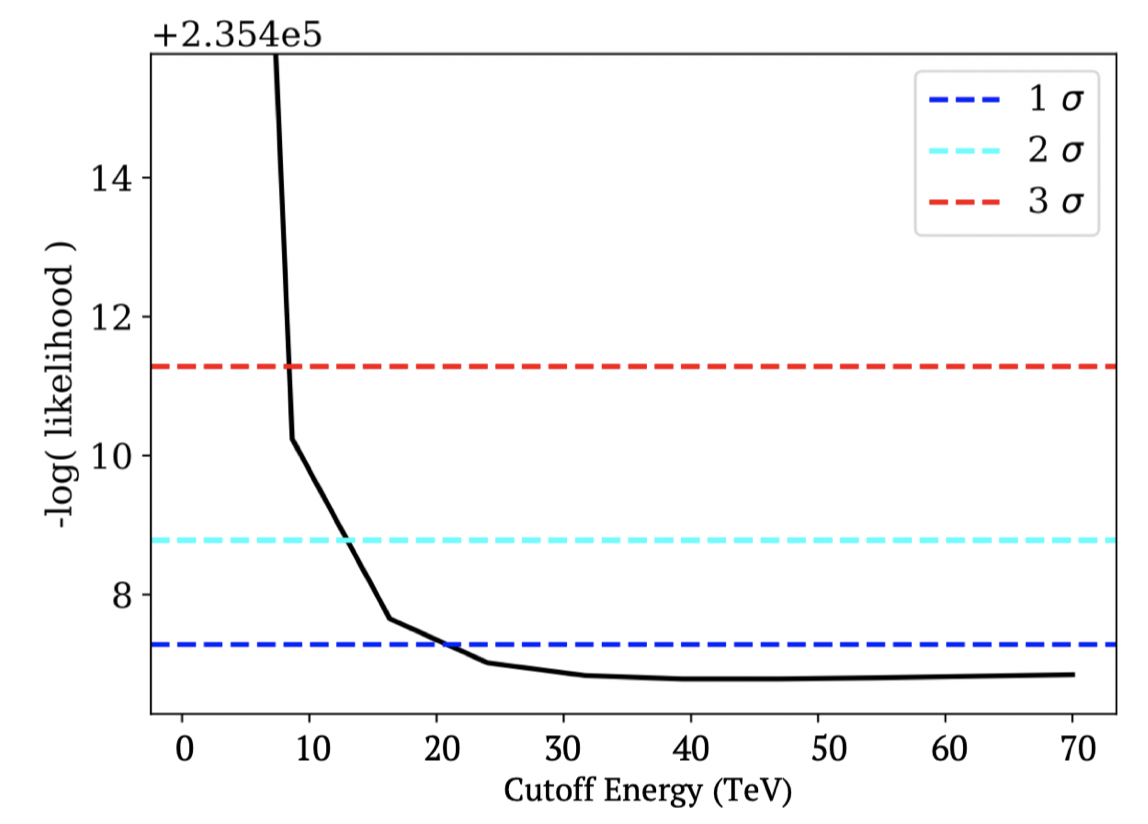}
 \end{subfigure}
 \begin{subfigure}{.5\textwidth}
  \includegraphics[width=0.9\textwidth]{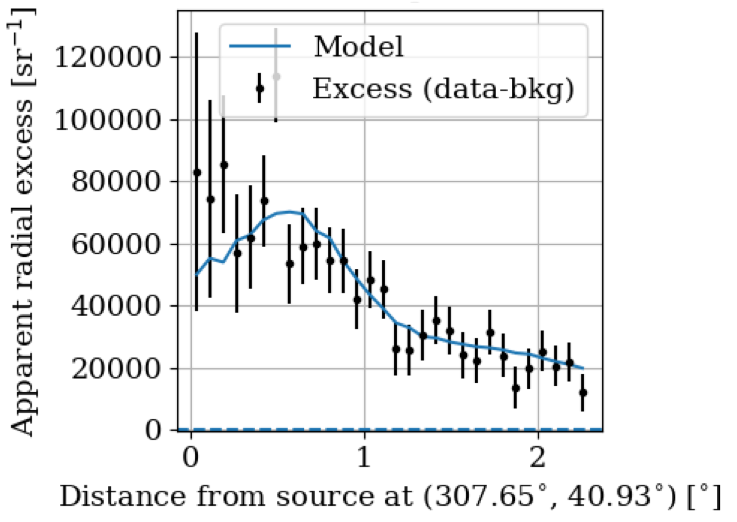}
 \end{subfigure}
 \captionsetup{width=.9\linewidth}
 \caption{{\label{fig:cocoon_cutoff_ll}} \textbf{Likelihood profile of the Cocoon cutoff energy and radial profile of the Cocoon region.} \textbf{Left (a)}: Likelihood profile of the cutoff energy for the Cygnus Cocoon assuming a power law spectrum with an exponential cutoff as in equation 3. \label{fig:radial_profile} \textbf{Right (b)}: Radial profile of the $\gamma$-ray emission at ROI region centered at the HAWC Cocoon location. The errors shown are 1 $\sigma$ statistical errors. The blue model curve is the PWN and $\gamma$ Cygni subtracted from the multi-source model in Extended Data Fig. 2. The initial bump in the blue curve is an artifact of the binning used.}
\end{figure}

\begin{figure}
\begin{subfigure}{.33\textwidth}
\captionsetup{width=.9\linewidth}
\includegraphics[scale=0.31]{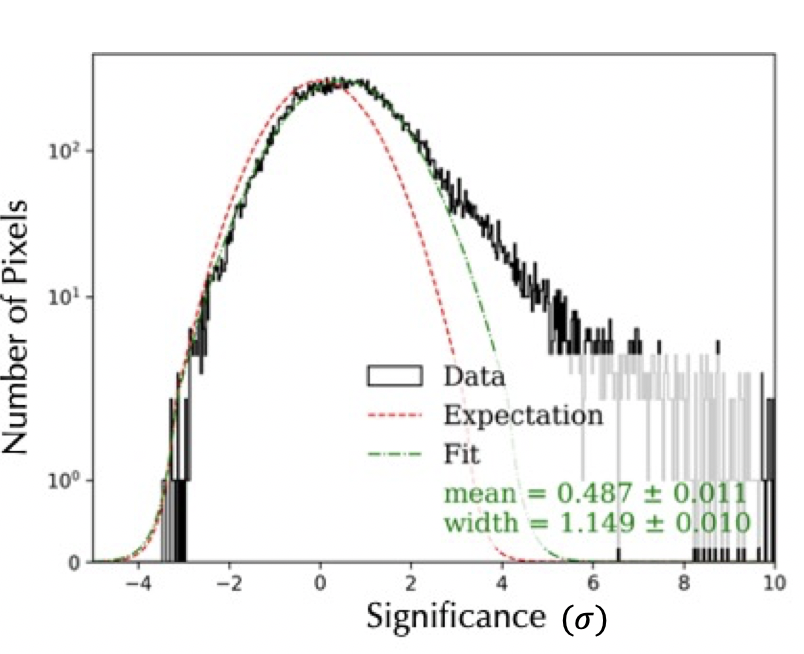}
\label{fig:whole_sig}
\end{subfigure}
\begin{subfigure}{.33\textwidth}
\captionsetup{width=.9\linewidth}
\includegraphics[scale=0.31]{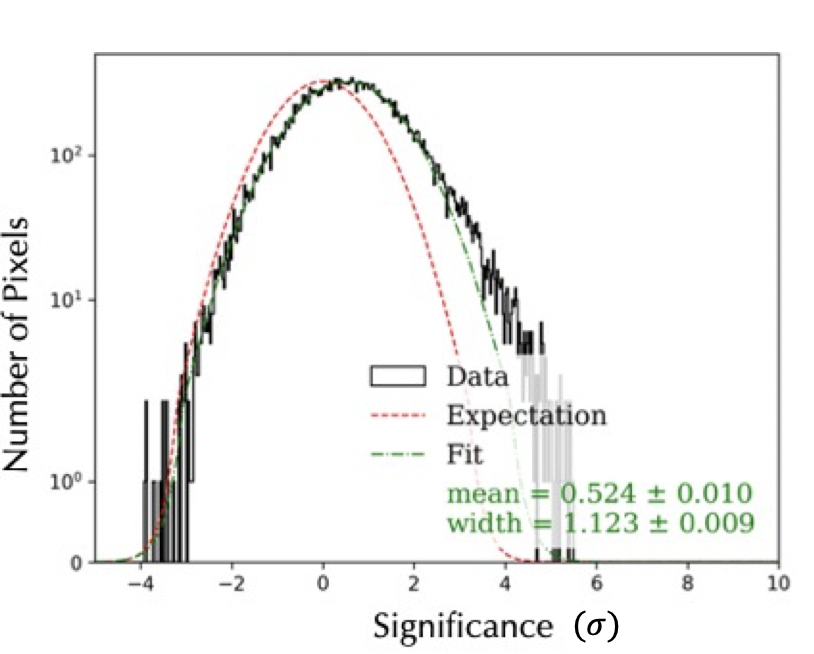}
\label{fig:sub1_sig}
\end{subfigure}
\begin{subfigure}{.33\textwidth}
\captionsetup{width=.9\linewidth}
\includegraphics[scale=0.31]{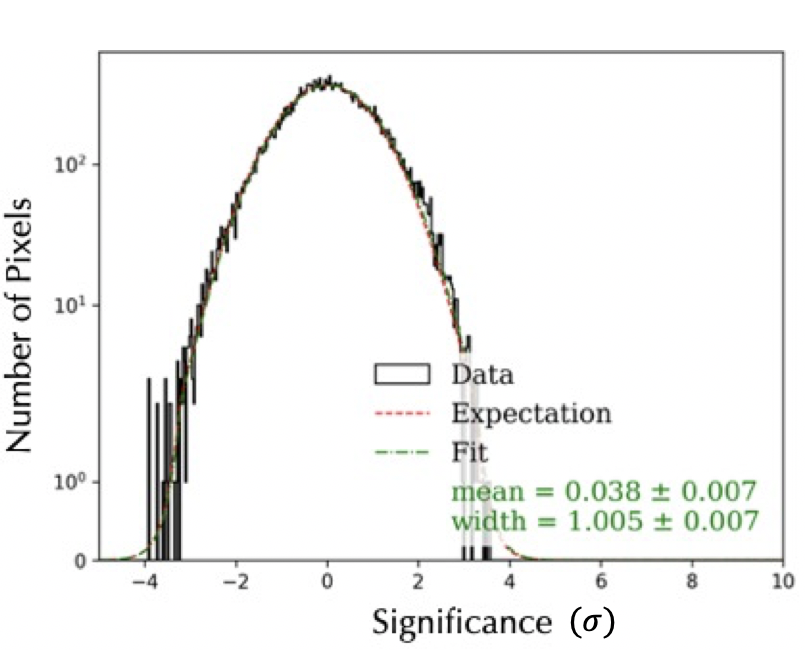}
\label{fig:sub2_sig}
\end{subfigure}
\captionsetup{width=.9\linewidth}
\caption{\textbf{Significance distribution in the analysis region of interest (ROI).}The black curve is the distribution of excess significance in the ROI. The red and the green curves are the expected and the obtained distribution respectively. If there is no significance above the background fluctuations, the significance histogram should follow a normal distribution where the mean of the distribution is close to zero and the width is about 1 $\sigma$. \textbf{Left (a)}: Distribution of excess significance before any source subtraction in the given ROI overlaid with the expected and the best fit Gaussian distribution. \textbf{Middle (b)}: Distribution of excess significance after subtracting PWN \& $\gamma$ Cygni. \textbf{Right (c)}: Distribution of excess significance after subtracting PWN \& $\gamma$ Cygni and Cocoon.}
\label{fig:three_sig_figures}
\end{figure}

\begin{figure} 
\begin{subfigure}{.5\textwidth}
 \includegraphics[width=0.9\textwidth]{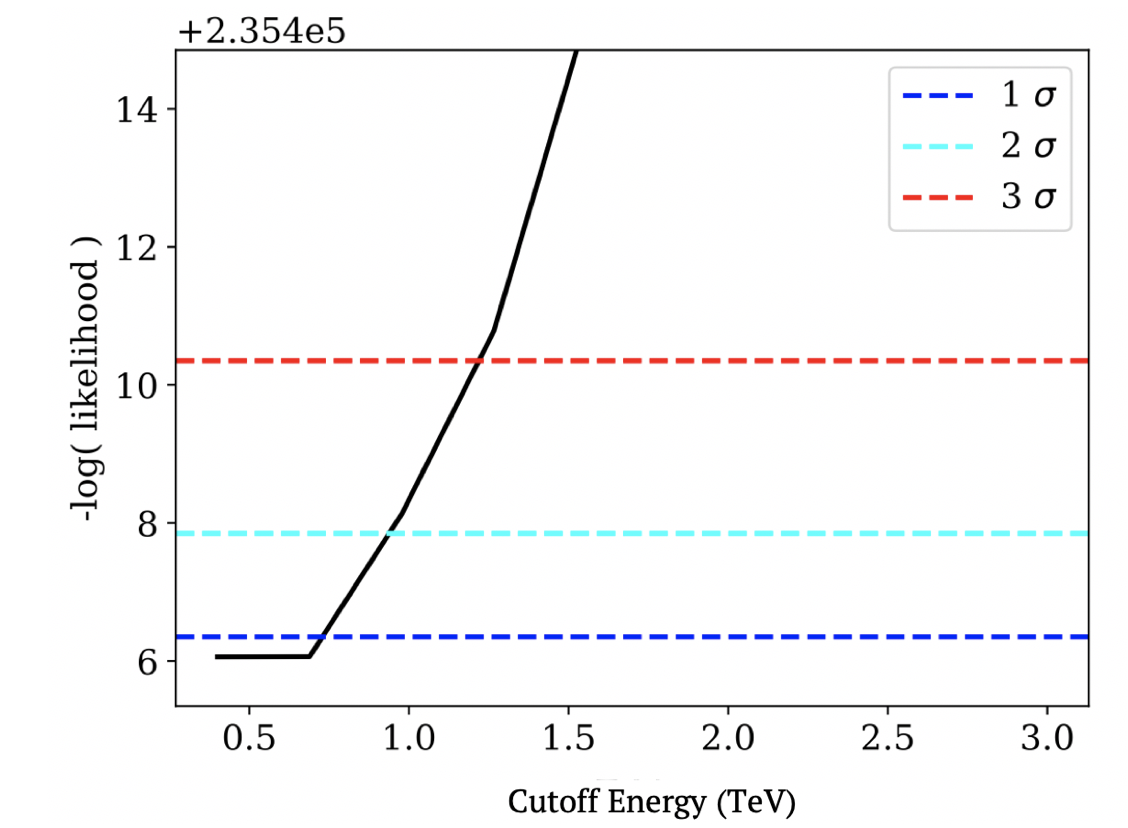}
 \end{subfigure}
 \begin{subfigure}{.5\textwidth}
  \includegraphics[width=0.9\textwidth]{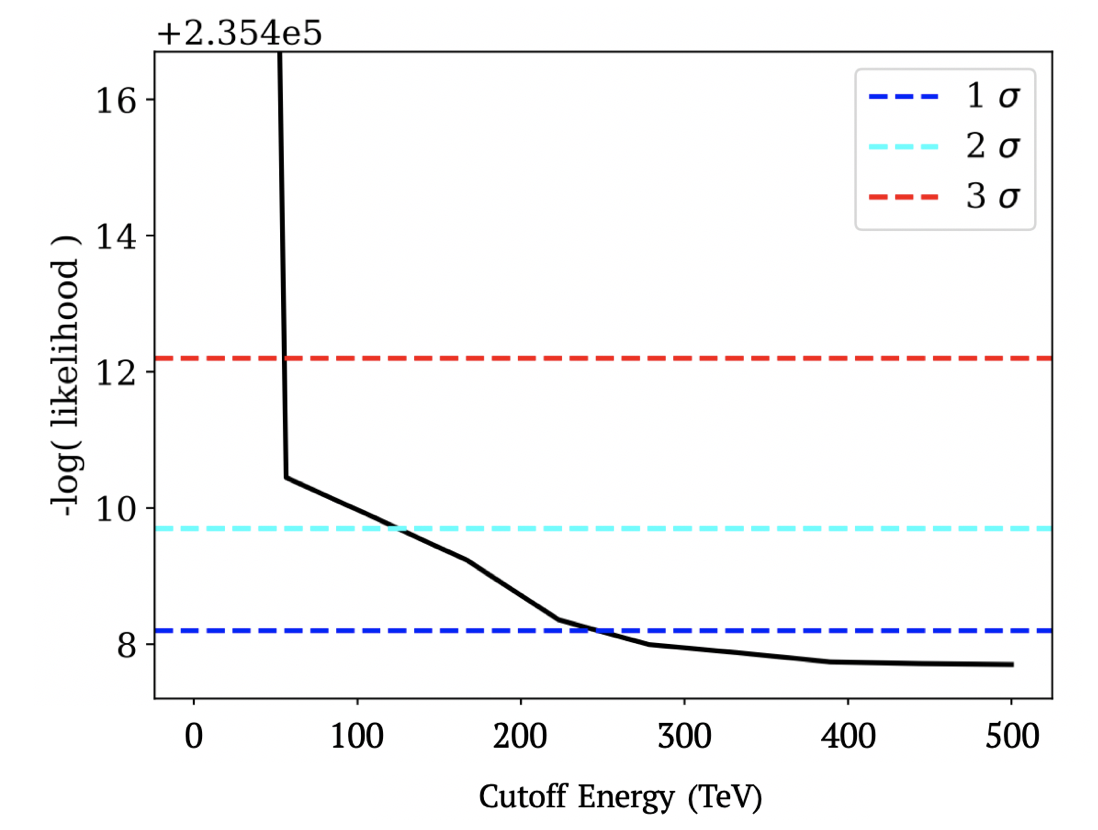}
 \end{subfigure}
 \captionsetup{width=.9\linewidth}
 \caption{\label{fig:energy_range} \textbf{Likelihood profile of the cutoff energy for the Cocoon energy range estimations} \textbf{Left}: Likelihood profile of hard cutoff to calculate the upper limit to the minimum Cocoon energy. \textbf{Right}: Likelihood profile of hard cutoff to calculate the lower limit to the maximum Cocoon energy.}
\end{figure}


\newpage

{\bf References}
\vspace{1em}

\begin{thebibliography}{99}
\expandafter\ifx\csname url\endcsname\relax
  \def\url#1{\texttt{#1}}\fi
\expandafter\ifx\csname urlprefix\endcsname\relax\def\urlprefix{URL }\fi
\providecommand{\bibinfo}[2]{#2}
\providecommand{\eprint}[2][]{\url{#2}}

\bibitem{hillasbook}
\bibinfo{author}{Hillas, A. M.}
\newblock \emph{\bibinfo{book}{Composition and Origin of Cosmic Rays}}.
\newblock
  (\bibinfo{year}{Dordrecht:Reidel, 1983}).
  

\bibitem{1990acr..book.....B}
\bibinfo{author}{{Berezinskii}, V.~S., {Bulanov}, S.~V., {Dogiel}, V.~A. \& {Ptuskin}, V.~S.}
\newblock \emph{\bibinfo{book}{Astrophysics of Cosmic Rays}}.
\newblock 
  (\bibinfo{year}{North Holland, 1990}).  
  
\bibitem{Baade259}
\bibinfo{author}{Baade, W. \& Zwicky, F.}
\newblock \bibinfo{title}{{Cosmic Rays from Super-Novae}}.
\newblock \emph{\bibinfo{journal}{Proc. Natl. Acad. Sci. U.S.A.}}
  \textbf{\bibinfo{volume}{20}}, \bibinfo{pages}{259--263}
  (\bibinfo{year}{1934}).  

\bibitem{H_randel_2004}
\bibinfo{author}{Hörandel, Jörg R.}
\newblock \bibinfo{title}{{Models of the knee in the energy spectrum of cosmic rays}}.
\newblock \emph{\bibinfo{journal}{Astropart. Phys.}}
  \textbf{\bibinfo{volume}{21}}, \bibinfo{pages}{241–265}
  (\bibinfo{year}{2004}).  
  
\bibitem{Bell:2013kq}
\bibinfo{author}{Bell, A. R., Schure, K. M., Reville, B. \&  Giacinti, G.}
\newblock \bibinfo{title}{{Cosmic ray acceleration and escape from supernova remnants}}.
\newblock \emph{\bibinfo{journal}{Mon. Not. Roy. Astron. Soc.}}
  \textbf{\bibinfo{volume}{431}}, \bibinfo{pages}{415}
  (\bibinfo{year}{2013}).  
  
\bibitem{2013APh....43...71A}
\bibinfo{author}{Aharonian, F.}
\newblock \bibinfo{title}{{Gamma rays from supernova remnants}}.
\newblock \emph{\bibinfo{journal}{Astropart. Phys.}}
  \textbf{\bibinfo{volume}{43}}, \bibinfo{pages}{71-80}
  (\bibinfo{year}{2013}).  
  
\bibitem{2012SSRv..173..369H}
\bibinfo{author}{Helder, E. A.} \emph{et~al.}
\newblock \bibinfo{title}{{Observational Signatures of Particle Acceleration in Supernova Remnants}}.
\newblock \emph{\bibinfo{journal}{Space Sci. Rev.}}
  \textbf{\bibinfo{volume}{173}}, \bibinfo{pages}{369-431}
  (\bibinfo{year}{2012}).  
  
\bibitem{2017hsn..book.1737F}
\bibinfo{author}{Funk, S.}
\newblock \bibinfo{title}{{High-Energy Gamma Rays from Supernova Remnants}}.
\newblock \emph{\bibinfo{book}{Handbook of Supernovae}}.
  (\bibinfo{year}{Springer, Cham, 2017}).  
  
\bibitem{2016Natur.531..476H}
\bibinfo{author}{Abramowski, A.} \emph{et~al.}
\newblock \bibinfo{title}{{Acceleration of petaelectronvolt protons in the Galactic Centre}}.
\newblock \emph{\bibinfo{journal}{Nature}}
  \textbf{\bibinfo{volume}{531}}, \bibinfo{pages}{476-479}
  (\bibinfo{year}{2016}).  
  
\bibitem{fermicocoon}
\bibinfo{author}{Ackermann, M.} \emph{et~al.}
\newblock \bibinfo{title}{{A Cocoon of Freshly Accelerated Cosmic Rays Detected by Fermi in the Cygnus Superbubble}}.
\newblock \emph{\bibinfo{journal}{Science}}
  \textbf{\bibinfo{volume}{334}}, \bibinfo{pages}{1103-1107}
  (\bibinfo{year}{2011}).  
  
\bibitem{catalog}
\bibinfo{author}{Abeysekara, A. U.} \emph{et~al.}
\newblock \bibinfo{title}{{The 2HWC HAWC Observatory Gamma-Ray Catalog}}.
\newblock \emph{\bibinfo{journal}{Astrophys. J.}}
  \textbf{\bibinfo{volume}{843}}, \bibinfo{pages}{40}
  (\bibinfo{year}{2017}).  
  
\bibitem{pwn2002}
\bibinfo{author}{Aharonian, F.} \emph{et~al.}
\newblock \bibinfo{title}{{An unidentified TeV source in the vicinity of Cygnus OB2}}.
\newblock \emph{\bibinfo{journal}{Astron. Astrophys.}}
  \textbf{\bibinfo{volume}{393}}, \bibinfo{pages}{L37-L40}
  (\bibinfo{year}{2002}).  
  
\bibitem{pwn2005}
\bibinfo{author}{Aharonian, F.} \emph{et~al.}
\newblock \bibinfo{title}{{The unidentified TeV source (TeV J2032+4130)  and surrounding field: Final HEGRA IACT-System results}}.
\newblock \emph{\bibinfo{journal}{Astron. Astrophys.}}
  \textbf{\bibinfo{volume}{431}}, \bibinfo{pages}{197-202}
  (\bibinfo{year}{2005}).    
  
\bibitem{Bartoli14}
\bibinfo{author}{Bartoli, B.} \emph{et~al.}
\newblock \bibinfo{title}{{Identification of the TeV Gamma-Ray Source ARGO J2031+4157 with the Cygnus Cocoon}}.
\newblock \emph{\bibinfo{journal}{Astrophys. J.}}
  \textbf{\bibinfo{volume}{790}}, \bibinfo{pages}{152}
  (\bibinfo{year}{2014}).  
  
\bibitem{4fgl}
\bibinfo{author}{Abdollahi, S.} \emph{et~al.}
\newblock \bibinfo{title}{{Fermi Large Area Telescope Fourth Source Catalog}}.
\newblock \emph{\bibinfo{journal}{Astrophys. J. Suppl.}}
  \textbf{\bibinfo{volume}{247}}, \bibinfo{pages}{33}
  (\bibinfo{year}{2020}).    

\bibitem{2003AJ....125.3145T}
\bibinfo{author}{Taylor, A. R.} \emph{et~al.}
\newblock \bibinfo{title}{{The Canadian Galactic Plane Survey}}.
\newblock \emph{\bibinfo{journal}{Astron. J.}}
  \textbf{\bibinfo{volume}{125}}, \bibinfo{pages}{3145–3164}
  (\bibinfo{year}{2003}).  
  
\bibitem{suzaku}
\bibinfo{author}{Mizuno, T.} \emph{et~al.}
\newblock \bibinfo{title}{{Suzaku Observation of the Fermi Cygnus Cocoon: The Search for a Signature of Young Cosmic-Ray Electrons}}.
\newblock \emph{\bibinfo{journal}{Astrophys. J.}}
  \textbf{\bibinfo{volume}{803}}, \bibinfo{pages}{74}
  (\bibinfo{year}{2015}).    


\bibitem{localCR_density}
\bibinfo{author}{Aguilar, M.} \emph{et~al.}
\newblock \bibinfo{title}{{Precision Measurement of the Proton Flux in Primary Cosmic Rays from Rigidity 1 GV to 1.8 TV with the Alpha Magnetic Spectrometer on the International Space Station}}.
\newblock \emph{\bibinfo{journal}{Phys. Rev. Lett.}}
  \textbf{\bibinfo{volume}{114}}, \bibinfo{pages}{171103}
  (\bibinfo{year}{2015}). 
  
\bibitem{aha_nature}
\bibinfo{author}{Aharonian, F., Yang, R. \& {de O{\~n}a Wilhelmi}, E.}
\newblock \bibinfo{title}{{Massive stars as major factories of Galactic cosmic rays}}.
\newblock \emph{\bibinfo{journal}{Nat. Astron.}}
  \textbf{\bibinfo{volume}{3}}, \bibinfo{pages}{561-567}
  (\bibinfo{year}{2019}).   


\bibitem{aharonian_book2}
\bibinfo{author}{Aharonian, F.}
\newblock \emph{\bibinfo{book}{Astrophysics at Very High Energies, Saas-Fee Advanced
Course}}.
  (\bibinfo{year}{Berlin, Heidelberg: Springer-Verlag, 2013}).  
  
\bibitem{agenew}
\bibinfo{author}{Wright, N. J. and Drew, J. E. \& Mohr-Smith, M.}
\newblock \bibinfo{title}{{The massive star population of Cygnus OB2}}.
\newblock \emph{\bibinfo{journal}{Mon. Not. Roy. Astron. Soc.}}
  \textbf{\bibinfo{volume}{449}}, \bibinfo{pages}{741-760}
  (\bibinfo{year}{2015}).  
  
\bibitem{butt}
\bibinfo{author}{Butt, Y.}
\newblock \bibinfo{title}{{Beyond the myth of the Supernovaremnant origin of Cosmic Rays}}.
\newblock \emph{\bibinfo{journal}{Nature}}
  \textbf{\bibinfo{volume}{460}}, \bibinfo{pages}{701}
  (\bibinfo{year}{2009}).  
  
\bibitem{OB2mass}
\bibinfo{author}{Kn{\"o}dlseder, J.}
\newblock \bibinfo{title}{{Cygnus OB2 - a young globular cluster in the Milky Way}}.
\newblock \emph{\bibinfo{journal}{Astron. Astrophys.}}
  \textbf{\bibinfo{volume}{360}}, \bibinfo{pages}{539-548}
  (\bibinfo{year}{2000}).  


\bibitem{OB2_2010}
\bibinfo{author}{Wright, N.J., Drake, J. J., Drew, J. E. \& Vink, J. S.}
\newblock \bibinfo{title}{{The Massive Star-Forming Region Cygnus OB2. II. Integrated Stellar Properties and the Star Formation History}}.
\newblock \emph{\bibinfo{journal}{Astrophys. J.}}
  \textbf{\bibinfo{volume}{713}}, \bibinfo{pages}{871-882}
  (\bibinfo{year}{2010}).  
  
\bibitem{icecube_ex}
\bibinfo{author}{Pinat, E. \& Sánchez, J. A. A.}
\newblock \bibinfo{title}{{Search for extended sources of neutrino emission with 7 years of IceCube data}}.
\newblock \emph{\bibinfo{proceedings}{PoS Proc. Sci.}}
  \textbf{\bibinfo{volume}{ICRC 2017}}, \bibinfo{pages}{963}
  (\bibinfo{year}{2018}).    


\bibitem{icecube_gen2}
\bibinfo{author}{Aartsen, M.G.} \emph{et~al.}
\newblock \bibinfo{title}{{IceCube-Gen2: The Window to the Extreme Universe}}.
\newblock \emph{\bibinfo{journal}{arXiv e-prints arXiv:2008.04323}}
  (\bibinfo{year}{2020}).  
  
\bibitem{swgo}
\bibinfo{author}{Schoorlemmer, H.}
\newblock \bibinfo{title}{{A next-generation ground-based wide field-of-view gamma-ray observatory in the southern hemisphere}}.
\newblock \emph{\bibinfo{journal}{36th International Cosmic Ray Conference (ICRC2019)}}
  \textbf{\bibinfo{volume}{36}}, \bibinfo{pages}{785}
  (\bibinfo{year}{2019}).  
  
\bibitem{lhaaso}
\bibinfo{author}{LHAASO telescope sensitivity to diffuse gamma-ray signals from the Galaxy}
\newblock \bibinfo{title}{{Neronov, A. and Semikoz, D.}}.
\newblock \emph{\bibinfo{journal}{Phys. Rev. D}}
  \textbf{\bibinfo{volume}{102}}, \bibinfo{pages}{043025}
  (\bibinfo{year}{2020}).   

\setcounter{firstbib}{\value{enumiv}}
\end{thebibliography}

\newpage

{\bf References}
\vspace{1em}

\begin{thebibliography}{99}
\setcounter{enumiv}{\value{firstbib}}

\expandafter\ifx\csname url\endcsname\relax
  \def\url#1{\texttt{#1}}\fi
\expandafter\ifx\csname urlprefix\endcsname\relax\def\urlprefix{URL }\fi
\providecommand{\bibinfo}[2]{#2}
\providecommand{\eprint}[2][]{\url{#2}}

\bibitem{hawccrab_new1}
\bibinfo{author}{Abeysekara, A. U.} \emph{et~al.}
\newblock \bibinfo{title}{{Measurement of the Crab Nebula Spectrum Past 100 TeV with HAWC}}.
\newblock \emph{\bibinfo{journal}{Astrophys. J.}}
  \textbf{\bibinfo{volume}{881}}, \bibinfo{pages}{134}
  (\bibinfo{year}{2019}).
  
\bibitem{Abey17crab1}
\bibinfo{author}{Abeysekara, A. U.} \emph{et~al.}
\newblock \bibinfo{title}{{Observation of the Crab Nebula with the HAWC Gamma-Ray Observatory}}.
\newblock \emph{\bibinfo{journal}{Astrophys. J.}}
  \textbf{\bibinfo{volume}{843}}, \bibinfo{pages}{39}
  (\bibinfo{year}{2017}).
  
\bibitem{3ml1}
\bibinfo{author}{Vianello, G.} \emph{et~al.}
\newblock \bibinfo{title}{{The Multi-Mission Maximum Likelihood framework (3ML)}}.
\newblock \emph{\bibinfo{journal}{PoS Proc. Sci.}}
  \textbf{\bibinfo{volume}{34th ICRC}}
  (\bibinfo{year}{2015}).  
  

\bibitem{HE_catalog1}
\bibinfo{author}{Abeysekara, A. U.} \emph{et~al.}
\newblock \bibinfo{title}{{Multiple Galactic Sources with Emission Above 56 TeV Detected by HAWC}}.
\newblock \emph{\bibinfo{journal}{Phys. Rev. Lett.}}
  \textbf{\bibinfo{volume}{124}}, \bibinfo{pages}{021102}
  (\bibinfo{year}{2020}).    

\bibitem{Hona:2019ysf1}
\bibinfo{author}{Hona, B. for the HAWC Collaboration.}
\newblock \bibinfo{title}{{Testing the Limits of Particle Acceleration in Cygnus OB2 with HAWC}}.
\newblock \emph{\bibinfo{journal}{PoS Proc. Sci.}}
  \textbf{\bibinfo{volume}{ICRC2019}}, \bibinfo{pages}{699}
  (\bibinfo{year}{2020}).  


\bibitem{whipple1}
\bibinfo{author}{Lang, M. J.} \emph{et~al.}
\newblock \bibinfo{title}{{Evidence for TeV gamma ray emission from TeV J2032+4130 in Whipple archival data}}.
\newblock \emph{\bibinfo{journal}{Astron. Astrophys.}}
  \textbf{\bibinfo{volume}{423}}, \bibinfo{pages}{415-419}
  (\bibinfo{year}{2004}).  
  
\bibitem{MAGIC1}
\bibinfo{author}{Albert, J.} \emph{et~al.}
\newblock \bibinfo{title}{{MAGIC Observations of the Unidentified {$\gamma$}-Ray Source TeV J2032+4130}}.
\newblock \emph{\bibinfo{journal}{Astrophys. J. Lett.}}
  \textbf{\bibinfo{volume}{675}}, \bibinfo{pages}{L25}
  (\bibinfo{year}{2008}).  
  
\bibitem{pwnmilagro1}
\bibinfo{author}{Abdo, A. A.} \emph{et~al.}
\newblock \bibinfo{title}{{Spectrum and Morphology of the Two Brightest Milagro Sources in the Cygnus Region: MGRO J2019+37 and MGRO J2031+41}}.
\newblock \emph{\bibinfo{journal}{Astrophys. J.}}
  \textbf{\bibinfo{volume}{753}}, \bibinfo{pages}{159}
  (\bibinfo{year}{2012}).  
  
\bibitem{veritas2014}
\bibinfo{author}{Aliu, E.} \emph{et~al.}
\newblock \bibinfo{title}{{Observations of the unidentified gamma-ray source TeV J2032+4130 by VERITAS}}.
\newblock \emph{\bibinfo{journal}{Astrophys. J.}}
  \textbf{\bibinfo{volume}{783}}, \bibinfo{pages}{16}
  (\bibinfo{year}{2014}).    
  
\bibitem{verj20191}
\bibinfo{author}{Aliu, E.} \emph{et~al.}
\newblock \bibinfo{title}{{Discovery of TeV Gamma-Ray Emission toward Supernova Remnant SNR G78.2+2.1}}.
\newblock \emph{\bibinfo{journal}{Astrophys. J.}}
  \textbf{\bibinfo{volume}{770}}, \bibinfo{pages}{93}
  (\bibinfo{year}{2013}).  
  
\bibitem{wilks1}
\bibinfo{author}{Wilks, S. S.}
\newblock \bibinfo{title}{{The Annals of Mathematical Statistics}}.
\newblock \emph{\bibinfo{journal}{Ann. Math. Stat.}}
  \textbf{\bibinfo{volume}{9}}, \bibinfo{pages}{60-62}
  (\bibinfo{year}{1938}). 
  
\bibitem{henrike1}
\bibinfo{author}{Fleischhack, H. for the HAWC Collaboration.}
\newblock \bibinfo{title}{{Modeling the non-thermal emission of the gamma Cygni Supernova Remnant up to the highest energies}}.
\newblock \emph{\bibinfo{journal}{PoS Proc. Sci.}}
  \textbf{\bibinfo{volume}{36th ICRC}}, \bibinfo{pages}{675}
  (\bibinfo{year}{2019}).  
  
\bibitem{geminga1}
\bibinfo{author}{Abeysekara, A. U.} \emph{et~al.}
\newblock \bibinfo{title}{{Extended gamma-ray sources around pulsars constrain the origin of the positron flux at Earth}}.
\newblock \emph{\bibinfo{journal}{Science}}
  \textbf{\bibinfo{volume}{358}}, \bibinfo{pages}{911-914}
  (\bibinfo{year}{2017}).  

\bibitem{ob2distance1}
\bibinfo{author}{Kn{\"o}dlseder, J.} \emph{et~al.}
\newblock \bibinfo{title}{{Gamma-ray line emission from OB associations and young open clusters. II. The Cygnus region}}.
\newblock \emph{\bibinfo{journal}{Astron. Astrophys.}}
  \textbf{\bibinfo{volume}{390}}, \bibinfo{pages}{945-960}
  (\bibinfo{year}{2002}).  
  
\bibitem{for_eta11}
\bibinfo{author}{Dermer, C.D.}
\newblock \bibinfo{title}{{Secondary production of neutral pi-mesons and the diffuse galactic gamma radiation}}.
\newblock \emph{\bibinfo{journal}{Astron. Astrophys.}}
  \textbf{\bibinfo{volume}{157}}, \bibinfo{pages}{223-229}
  (\bibinfo{year}{1986}).  
  
\bibitem{for_eta21}
\bibinfo{author}{Kafexhiu, E., Aharonian, F., Taylor, Andrew M. \& Vila, G. S.}
\newblock \bibinfo{title}{{Parametrization of gamma-ray production cross sections for $pp$ interactions in a broad proton energy range from the kinematic threshold to PeV energies}}.
\newblock \emph{\bibinfo{journal}{Phys. Rev. D}}
  \textbf{\bibinfo{volume}{90}}, \bibinfo{pages}{123014}
  (\bibinfo{year}{2014}).  
  
\bibitem{kelner1}
\bibinfo{author}{Kelner, S. R., Aharonian, F. A. \& Bugayov, V. V.}
\newblock \bibinfo{title}{{Energy spectra of gamma rays, electrons, and neutrinos produced at proton-proton interactions in the very high energy regime}}.
\newblock \emph{\bibinfo{journal}{Phys. Rev. D}}
  \textbf{\bibinfo{volume}{74}}, \bibinfo{pages}{034018}
  (\bibinfo{year}{2006}).  
  
\bibitem{burstsolution1}
\bibinfo{author}{Syrovatskii, S. I.}
\newblock \bibinfo{title}{{The Distribution of Relativistic Electrons in the Galaxy and the Spectrum of Synchrotron Radio Emission.}}.
\newblock \emph{\bibinfo{journal}{Sov. Astron.}}
  \textbf{\bibinfo{volume}{3}}, \bibinfo{pages}{22}
  (\bibinfo{year}{1959}).  
  
\bibitem{drury1}
\bibinfo{author}{Drury, L. O'C.}
\newblock \bibinfo{title}{{Origin of cosmic rays}}.
\newblock \emph{\bibinfo{journal}{Astropart. Phys.}}
  \textbf{\bibinfo{volume}{39}}, \bibinfo{pages}{52-60}
  (\bibinfo{year}{2012}).  
  
\bibitem{oblist11}
\bibinfo{author}{Mel'Nik, A. M. \& Efremov, Y. N.}
\newblock \bibinfo{title}{{A new list of OB associations in our galaxy}}.
\newblock \emph{\bibinfo{journal}{Astron. Lett.}}
  \textbf{\bibinfo{volume}{21}}, \bibinfo{pages}{10-26}
  (\bibinfo{year}{1995}).  

\bibitem{oblist21}
\bibinfo{author}{Kharchenko, N. V., Piskunov, A. E., Schilbach, E., R{\"o}ser, S. \& Scholz, R. D.}
\newblock \bibinfo{title}{{Global survey of star clusters in the Milky Way. II. The catalogue of basic parameters}}.
\newblock \emph{\bibinfo{journal}{Astron. Astrophys.}}
  \textbf{\bibinfo{volume}{558}}, \bibinfo{pages}{A53}
  (\bibinfo{year}{2013}). 
  
\bibitem{2015PhRvD..92h3003P1}
\bibinfo{author}{Prosekin, A. Y., Kelner, S. R. \& Aharonian, F. A.}
\newblock \bibinfo{title}{{Transition of propagation of relativistic particles from the ballistic to the diffusion regime}}.
\newblock \emph{\bibinfo{journal}{Phys. Rev. D}}
  \textbf{\bibinfo{volume}{92}}, \bibinfo{pages}{083003}
  (\bibinfo{year}{2015}).   
  
\bibitem{3fhl1}
\bibinfo{author}{Ajello, M.} \emph{et~al.}
\newblock \bibinfo{title}{{3FHL: The Third Catalog of Hard Fermi-LAT Sources}}.
\newblock \emph{\bibinfo{journal}{Astrophys. J., Suppl. Ser.}}
  \textbf{\bibinfo{volume}{232}}, \bibinfo{pages}{18}
  (\bibinfo{year}{2017}).    
  
\end{thebibliography}
\end{document}